\documentclass[]{aa}
\usepackage{amssymb}
\usepackage{amsmath}
\usepackage[varg]{txfonts}
\usepackage{multirow}
\usepackage[boxed,linesnumbered]{algorithm2e}
\usepackage{hyperref}
\usepackage{graphicx}

\usepackage{verbatim}
\usepackage{natbib}
\usepackage{rotate}
\usepackage{lscape}
\usepackage{aalongtable}
\usepackage{supertabular}
\usepackage{mathbbol}
\usepackage{bm}
\usepackage{mathtools}
\usepackage{enumerate}
\usepackage{booktabs}

\let\oldsqrt\sqrt
\def\sqrt{\mathpalette\DHLhksqrt}
\def\DHLhksqrt#1#2{%
\setbox0=\hbox{$#1\oldsqrt{#2\,}$}\dimen0=\ht0
\advance\dimen0-0.2\ht0
\setbox2=\hbox{\vrule height\ht0 depth -\dimen0}%
{\box0\lower0.4pt\box2}}

\def\aap{A\&A}

\def\u{{u}}
\def\v{{v}}
\def\w{{w}}
\def\l{{l}}
\def\m{{m}}
\def\n{{n}}

\def\RIME/{{\sc rime}}
\def\CLEAN/{{\sc clean}}
\def\Sec/{Sect.}
\def\MTMSCLEAN/{{\sc mtms-clean}}
\def\SSDCLEAN/{{\sc ssd-clean}}
\def\WSCLEAN/{{\sc ssd-clean}}
\def\DDFacet/{{\sc ddf}acet}
\def\SSD/{{\sc ssd}}
\def\BDA/{{\sc bda}}
\def\SSDGA/{\SSD/{\sc ga}}
\def\PMP/{{\sc hmp}}
\def\FW/{{\sc fw}}

\def\Montblanc/{{\sc montblanc}}
\newcommand{\sinc}[1]{\textrm{sinc}\left\{#1\right\}}

\newcommand{\Abs}[1]{\|#1\|}
\newcommand{\Cov}[1]{\textrm{Cov}\{#1\}}

\newcommand{\argmax}[1]{\underset{#1}{\mathrm{argmax}}}
\newcommand{\argmin}[1]{\underset{#1}{\mathrm{argmin}}}
\newcommand{\KER}[1]{\textrm{ker}\left\{#1\right\}}

\def\OmegaB{\Omega_\nu}

\def\npix{n_\mathrm{x}}
\def\npar{n_\mathrm{p}}
\def\nvis{n_\mathrm{v}}

\def\vecxnu{\vec{x}_{\nu}}
\def\vecynu{\vec{y}_{\nu}}
\def\vecbnu{\vec{b}_{\nu}}
\def\vecs{\vec{s}}
\def\vecb{\vec{b}}
\def\vecbnu{\vecb_\nu}
\def\BC{{\vecbnu}}

\def\DirtyVec{\vec{y}}
\def\DirtyVecNorm{\widetilde{\vec{y}}}
\def\DirtyVecNormChan{\widetilde{\vec{y}_{\nu}}}
\def\DirtyVecChan{\vec{y}_{\nu}}

\def\DirtyVecNormPixel{\widetilde{\vec{y}}_{i}}

\def\DirtyVecNormChanPixel{\widetilde{\vec{y}}_{i,\nu}}

\def\DeltaDirtyVecNorm{\widetilde{\vec{\delta y}}}
\def\DeltaDirtyVec{\vec{\delta y}}

\def\kron{\otimes}

\newcommand{\simeqdefT}[2][1.5]{
  \mathrel{\overset{\mathrm{def #2}}{\scalebox{#1}[1]{$\approx$}}}
}

\newcommand{\eqdef}{\overset{\mathrm{def}}{=\joinrel=}}
\newcommand{\eqdefT}[1]{\overset{\mathrm{def #1}}{=\joinrel=\joinrel=}}
\newcommand{\longeqdefT}[1]{\overset{\mathrm{def #1}}{=\joinrel=\joinrel=\joinrel=\joinrel=}}
\newcommand{\ddt}[1]{\frac{\textrm{d}{#1}}{\textrm{dt}}}

\newcommand{\smallM}{{\scriptstyle \bm{\mathcal{M}}}}

\def\JonesMat{\bm{\mathrm{G}}}

\def\MatFacetCoord{\bm{\mathrm{C}}_{\vec{s}_{\varphi}}}

\def\Unity{\bm{\mathrm{I}}}
\def\Fourier{\bm{\mathcal{F}}}
\def\MuellerImage{{\bm{\mathcal{M}}_{\nu}}}
\def\MuellerImageBL{{\bm{\mathcal{M}}_{\BC}}}
\def\MuellerImageBLH{{\bm{\mathcal{M}}^H_{\BC}}}
\def\MuellerImageDir{{\mathbf{M}^{\BC}}}
\newcommand{\MuellerImageDirBL}[1]{{\mathbf{M}^{\BC}_{#1}}}
\def\MuellerNorm{\widetilde{\MuellerImage}}

\def\TransferFunc{{\bm{\mathcal{T}}_{\nu}}}
\def\TransferFuncNorm{\widetilde{\TransferFunc}}
\def\TransferFuncDir{{\bm{\mathcal{T}}_{\nu}}}
\def\TransferFuncDirNorm{\widetilde{\TransferFuncDir}}

\def\SamplingBL{\bm{\mathcal{S}}_{\BC}}

\def\VisBL{\bm{\mathrm{v}}_{\BC}}
\def\ResVisBL{\bm{\mathrm{r}}_{\BC}}
\def\Vis{\bm{\mathrm{v}}}
\def\VisChan{\bm{\mathrm{v}}_\nu}

\def\WeightNu{\bm{\mathcal{W}}_{\nu}}
\def\PSFMatBL{\bm{\mathcal{C}}_{\BC}}

\def\EstimatedSky{\widehat{\vec{x_{\varphi}}}}
\def\MuellerImageBLConst{\MuellerImageBL^{\varphi}}
\def\MuellerImageBLConstT{{\MuellerImageBL^{\varphi}}^H}
\def\MuellerImageConst{\MuellerImage^{\varphi}}

\def\piAtPixel{\vec{\pi}_{i}}
\def\TransferFuncPhiPi{{\widetilde{\bm{\Theta}_{i}}}}
\def\TransferFuncPhi{\widetilde{\bm{\Theta}}}
\def\TransferFuncPhiNoNorm{\bm{\Theta}}
\def\piAtIsland{\vec{\pi_{\scriptscriptstyle\mathcal{I}}}}
\def\piGlobal{\vec{\pi}}
\def\PiLocalChan{{\bm{\Pi}^i_\nu}}

\def\TransferFuncPhiPiIsland{\bm{\Theta_{\bm{\mathcal{\scriptscriptstyle
          I}}}}}

\def\L{\bm{\mathrm{L}}}

\newcommand{\OMSOK}[1]{}

\usepackage{verbatim}
\usepackage{alltt}
\usepackage{upgreek}
\usepackage{soul}

\authorrunning{C. Tasse}

\title{Faceting for direction-dependent spectral deconvolution}

\author{C. Tasse\inst{1,2}, B. Hugo\inst{2,3}, M. Mirmont,
  O. Smirnov\inst{2,3}, M. Atemkeng\inst{2}, L. Bester\inst{3}, 
  M.J. Hardcastle\inst{4},
  R. Lakhoo\inst{5,6},
S. Perkins\inst{3} 
\and T. Shimwell\inst{7}}

\institute{GEPI, Observatoire de Paris, CNRS, Universit\'e Paris Diderot,
5 place Jules Janssen, 92190 Meudon, France
\and
Department of Physics \& Electronics, Rhodes University, PO Box 94,
Grahamstown, 6140, South Africa
\and
SKA South Africa, 3rd Floor, The Park, Park Road, Pinelands, 7405, South Africa
\and
Centre for Astrophysics Research, School of Physics, Astronomy and
Mathematics, University of Hertfordshire, College Lane, Hatfield AL10
9AB, UK
\and
Oxford e-Research Centre, University of Oxford, 7 Keble Road, Oxford, OX1 3QG
\and
Wolfson College, University of Oxford, Linton Road, Oxford, OX2 6UD
\and
Leiden Observatory, Leiden University, PO Box 9513, 2300 RA Leiden, the Netherlands
}

\abstract{The new generation of radio interferometers is characterized by high
sensitivity, wide fields of view and large fractional bandwidth. To
synthesize the deepest images enabled by the high dynamic range of
these instruments requires us to take into account
the direction-dependent Jones matrices, while estimating
the spectral properties of the sky in the imaging and deconvolution
algorithms. 

In this paper we discuss and implement a wideband wide-field direction-dependent spectral
deconvolution framework (\DDFacet/) based on image plane faceting,
that takes into account generic direction-dependent effects. Specifically, we present a
wide-field co-planar faceting scheme, and discuss the various effects
that need to be taken into account to solve for the deconvolution
problem (image plane normalization, position-dependent Point Spread Function, etc). We discuss two wideband spectral deconvolution
algorithms based on hybrid matching
pursuit and sub-space optimisation respectively. A few interesting technical features incorporated in our
imager are discussed, including baseline dependent averaging, which
has the effect of improving computing efficiency. The version of \DDFacet/ presented here 
can account for any externally defined Jones matrices and/or
beam patterns.

}

\date{}

\begin{document}

\maketitle


\section{Introduction}
\label{sec:Intro}

The new generation of interferometers is characterized by very wide
fields of view, large fractional bandwidth, high sensitivity, and high
resolution. The cross-correlation between voltages from pairs of
antenna (the visibilities) are often affected by severe 
baseline-time-frequency direction-dependent effects (DDEs) such as the
complex beam patterns (pointing errors, dish deformation, antenna
coupling within phased arrays), or by the ionosphere and its
associated Faraday rotation. The dynamic range needed to achieve the deepest
extragalactic surveys involves calibrating for DDEs \citep[see][and
  references
  therein]{Noordam10,Kazemi11,Tasse14a,Tasse14b,Smirnov15} and
taking them into account
in the imaging and deconvolution algorithms. 

The present paper discusses the
issues associated with estimating spatial and spectral
properties of the sky in the presence of DDEs. Those can be taken into account (i) in the Fourier domain
using A-Projection \citep{Bhatnagar08,Tasse13}, or (ii) in the image
domain using a facet approach
\citep{Weeren16,Williams16}. Algorithms of type (i) have the advantage
of giving a smooth image plane correction, while (ii) can give rise to
various discontinuity effects. However, (i) is often impractical in
the framework of DDE calibration, since a continuous (analytic) image plane
description of the Jones matrices has to be provided, while most
calibration schemes estimate Jones matrices in a discrete set of
directions. An additional step would be to spatially interpolate the
DDE calibration solutions, but this often proves to be difficult
due to the very nature of the Jones matrices ($2\times2$ complex valued), and
to the unitary ambiguity \citep[see][for a discussion on estimating
  beam pattern from sets of direction-dependent Jones matrices]{Yatawatta13}. Instead,
in this paper we address the issue of imaging and deconvolution in the
presence of generic DDEs using the faceting approach. 

In
\Sec/ \ref{sec:STPFacetting}, we present a general method of imaging
non-coplanar data on a multi-facet single tangential plane using
modified W-projection kernels (\FW/ kernels). This is a generalization
of the original idea presented by \citet{Kogan09}. In
\Sec/ \ref{sec:DDEFacetting}, we describe the non-trivial effects that
arise when forming a dirty image from a set of visibilities and Jones
matrices. Specifically, the vast majority of modern interferometers have large
fractional bandwidth and, since the station (or antenna\footnote{Throughout the paper,
we use the terms {stations} and {antennas} interchangably to
refer to the elements of an interferometer.}) beams scale with frequency, the 
effective PSF varies across the field of
view. We therefore stress here that even if (i) the effect of
decorrelation is minimized, and (ii) DDEs are corrected for, 
all existing imagers will give biased morphological results (unresolved sources will appear to
be larger toward the edge of the field, since higher-order spatial frequencies are 
preferentially attenuated by the beam response). 
The imaging and deconvolution framework presented here  (\DDFacet/\footnote{\DDFacet/ is an
  imager, implemented in C and Python. It
  will be made publicly available shortly after the publication
  of this paper.}) is
the first to take all these combined effects into account. 

Another important aspect of the work presented in this paper is
wide-band spectral deconvolution: the large fractional bandwidth of modern radio interferometers and the
need for high dynamic range images means deconvolution algorithms need
to estimate the intrinsic continuum sources spectral properties. This is
routinely done by the widely used wide band \MTMSCLEAN/ 
deconvolution algorithm \citep{Rau11}. An alternative approach has been implemented
by \citet{wsclean} in the WSCLEAN package. However, since the sources are affected by
frequency dependent DDEs, one needs to couple wide-band and DDE
deconvolution algorithms. To our knowledge only \citet{Bhatnagar13}
have implemented such an algorithm, but it is heavily reliant on the
assumption that the antennas are dishes, and most directly applicable to the VLA. Also, it
does not deal with baseline-time-frequency dependent DDEs (which give rise
to a direction-dependent PSF). We aim to build a flexible framework that can solve the wide-band
deconvolution problem in the presence of generic DDEs. Specifically, in
\Sec/ \ref{sec:WBDeconv}, we
present two wide-band deconvolution algorithms that natively
incorporate and compensate for the DDE effects discussed above. The
first uses a variation of a matching pursuit algorithm to which we
have added an optimisation step. The second uses joint deconvolution
on subsets of pixels, and we refer to it as a subspace
deconvolution. We have implemented one such approach using a genetic
algorithm.

In \Sec/ \ref{sec:Implementation}, we present an implementation
of the
ideas presented in this paper. Our implementation includes baseline-dependent
averaging (\Sec/ \ref{sec:BDA}), and irregular tessellation
(\Sec/ \ref{sec:tessel}). A simulation is discussed in detail in
\Sec/ \ref{sec:Simulation}. We outline the main results of this paper in
\Sec/ \ref{sec:Conclusion}.

\section{Faceting for wide field imaging}
\label{sec:STPFacetting}


The purpose of faceting is to approximate a wider field of view with many small narrow-field
images. \citet{Cornwell92} have proposed a polyhedron-like faceting
approach, where each narrow-field facet is tangent to the celestial
sphere at its own phase center. One of the biggest drawbacks of the noncoplanar polyhedron faceting
approach is that the minor cycle deconvolution becomes
complicated. Specifically, one needs to re-project each noncoplanar
facet into a single plane after synthesis (i.e. in the
image-space). Doing the necessary re-projections and inevitable (and
expensive) corrections for the areas where the facets overlap can be
done through astronomy mosaicing software packages such as the
Montage suite \citep{Jacob04}.

Alternatively, \citet{Kogan09} have described a fundamentally
different technique allowing one to build the facets onto a single, common
tangential plane. This algorithm consists in applying $\w$-dependent
$(\u,\v)$ coordinate transformation. However, phase errors due to noncoplanarity
quickly become a problem, and a W-projection type \citep{Cornwell08}
correction needs to be applied. As shown in
\Sec/ \ref{sec:AccurateFacetting} the kernels we are applying are
facet-dependent, and differ from the classical W-projection kernels
(the gridding and degridding algorithms are described in detail in
\Sec/ \ref{sec:BDA}).

\subsection{Formalism for the faceting}
\label{sec:RIME}


In order to model the complex direction-dependent effects (DDE - station beams,
ionosphere, Faraday rotation, etc), we use the
Radio Interferometry Measurement Equation (RIME) formalism, which
provides a model of a generic interferometer \citep[for
extensive discussions of the validity and limitations of the measurement
equation see][]{Hamaker96,Oleg11}. Each of the physical
phenomena that transform or convert the electric field before the
correlation is modeled by linear transformations
(2$\times$2 matrices). If $\vec{s}=[\l,\m,\n=\sqrt{1-\l^2-\m^2}]^T$ is a sky
direction, and $\textbf{M}^{H}$ stands for the Hermitian transpose operator of
matrix $\textbf{M}$,
then the $2\times 2$ correlation matrix $\textbf{V}_{(pq)t\nu}$ between antennas
$p$ and $q$ at time $t$ and frequency $\nu$ can
be written as:

\begin{alignat}{2}
\label{eq:ME}
\textbf{V}_{(pq),t\nu}=& 
\int_{\vec{s}}
\left(\JonesMat_{p\vec{s}t\nu}\textbf{X}_{\vec{s}}\ \JonesMat^H_{q\vec{s}t\nu}\right)
k^{\vec{s}}_{(pq),t\nu} \textrm{d}\vec{s}+n_{(pq),t\nu}
\\
\label{eq:kterm}
\text{with }k^{\vec{s}}_{(pq),t\nu}=&\exp{\left(-2 i\pi \frac{\nu}{c}\left(\vec{b}_{pq,t}^T(\vec{s}-\vec{s}_0)\right)\right)}\\
\text{and }\vec{b}_{pq,t}=&
\begin{bmatrix}
\u_{pq,t} \\ 
\v_{pq,t} \\ 
\w_{pq,t} 
\end{bmatrix}=
\begin{bmatrix}
\u_{p,t} \\ 
\v_{p,t} \\ 
\w_{p,t} 
\end{bmatrix}-
\begin{bmatrix}
\u_{q,t} \\ 
\v_{q,t} \\ 
\w_{q,t} 
\end{bmatrix}\\
\text{and }
\vec{s}=&
\begin{bmatrix}
\l \\ 
\m \\ 
\n 
\end{bmatrix}
\text{ and }\vec{s}_0=
\begin{bmatrix}
0 \\ 
0 \\ 
1 
\end{bmatrix}
\end{alignat}

\def\EXp{\textrm{E}_{X}}
\def\EYp{\textrm{E}_{Y}}
\def\EXq{\textrm{E}_{X}}
\def\EYq{\textrm{E}_{Y}}

\noindent where $\vec{b}_{pq,t}$ is the baseline
vector between antennas $p$ and $q$. The scalar term $k^{\vec{s}}_{(pq)t\nu}$ describes the
effect of the array geometry and correlator on the observed phase
shift of a coherent plane wave between antennas $p$ and $q$.
The $2\times2$ matrix $\JonesMat_{p\vec{s}t\nu}$ is the product of direction-dependent
Jones matrices corresponding to antenna $p$ (e.g., beam, ionosphere
phase screen and Faraday rotation), and 
$\textbf{X}_{\vec{s}}$ is
referred as the {\it sky
  term}
in the direction $\vec{s}$, and is the true
underlying source coherency matrix. The term 
$n_{(pq),t\nu}$ is a random variable modeling the system noise. In
the following however, we
will assume $\textrm{E}\left\{n_{(pq),t\nu}\right\}=0$ and implicitly work on the expected values
$\textrm{E}\left\{.\right\}$ rather than on the random
variables (except in \Sec/ \ref{sec:LossCovProp} and \ref{sec:DDEFacetting_IMCORR}, where we
describe the structure of the noise in the image domain).
Making the $(t\nu)$ indices implicit, we can transform Eq. \ref{eq:kterm} to:

\begin{alignat}{2}
\label{eq:facet}
\textbf{V}_{pq}=&
\displaystyle\sum\limits_{\varphi} 
\left[
\int_{\vec{s}\in\Omega_\varphi}
\left(\JonesMat_p\textbf{X}_{\vec{s}}\ \JonesMat_q^H\right)
k^{\vec{s}}_{pq,\varphi}
\right]
\\
k^{\vec{s}}_{pq,\varphi}=&
\exp{
\left(-2 i\pi \frac{\nu}{c}
\vec{b}_{pq}^T
\left(\vec{s}_\varphi-\vec{s}_0\right)\right)}\label{eq:PhaseFacet}\\
&
\exp{
\left(-2 i\pi \frac{\nu}{c}
\vec{b}^T_{pq}
\vec{\delta s}_\varphi\right)}\label{eq:KernelFacet}
\end{alignat}


\noindent where $\vec{s}_\varphi=[\l_{\varphi},\m_{\varphi},\n_{\varphi}]^{T}$ is the direction of the facet phase center and
$\bm{\delta} \vec{s}_\varphi=\vec{s}-\vec{s}_\varphi=
[\l-\l_{\varphi},
\m-\m_{\varphi},
\n-\n_{\varphi}]
=
[\delta \l_\varphi,
\delta \m_\varphi,
\delta \n_\varphi
]$ are the sky coordinates in the reference frame
of facet $\varphi$.

Applying the term \ref{eq:PhaseFacet} in
Eq. \ref{eq:PhaseFacet}+\ref{eq:KernelFacet} to the visibilities, one
still need to apply the position dependent term \ref{eq:KernelFacet},
which can be decomposed as:

\begin{alignat}{2}
\label{eq:facet}
\exp{
\left(-2 i\pi \frac{\nu}{c}
\vec{b}_{pq}^T
\vec{\delta s}_\varphi\right)}=&
\exp{
\left(
-2 i\pi \frac{\nu}{c}
(\u\delta \l_\varphi+
\v\delta \m_\varphi)
\right)}\\
&\exp{
\left(-2 i\pi \frac{\nu}{c}
\w\delta \n_\varphi)
\right)}
\end{alignat}



The second exponential term is the analog of the $w$-term corrected for in the W-projection style
algorithms. As pointed out by \citet{Kogan09}, the first order Taylor expansion approximation of $\delta n_\varphi$ can be written as:

\begin{alignat}{2}
\label{eq:dni}
\delta n_\varphi\approx&-\frac{1}{\sqrt{1-\l_\varphi^2-\m_\varphi^2}}\left(
\l_\varphi\delta \l_\varphi+\m_\varphi\delta \m_\varphi
\right)
\end{alignat}

\noindent which, conveniently, can be expressed in terms of a coordinate change in $uv$:

\begin{alignat}{2}
\label{eq:Kogan}
\exp{
\left(-2 i\pi \frac{\nu}{c}
\vec{b}_{pq}^T
\vec{\delta s}_\varphi\right)}
\approx&
\exp{
\left(-2 i\pi \frac{\nu}{c}
(\u'\delta \l_\varphi+
\v'\delta \m_\varphi)
\right)}
\end{alignat}

\noindent with $\u'=\u-\w\frac{\l_\varphi}{\sqrt{1-\l_\varphi^2-\m_\varphi^2}}$ and 
$\v'=\v-\w\frac{\m_\varphi}{\sqrt{1-\l_\varphi^2-\m_\varphi^2}}$. The linear approximation of $\delta n_i$ (Eq. \ref{eq:dni}) is plotted in
Fig. \ref{fig:Comparison}.

\subsection{Accurate noncoplanar faceting}
\label{sec:AccurateFacetting}

\begin{figure}[t!]
\begin{center}
\includegraphics[width=\columnwidth]{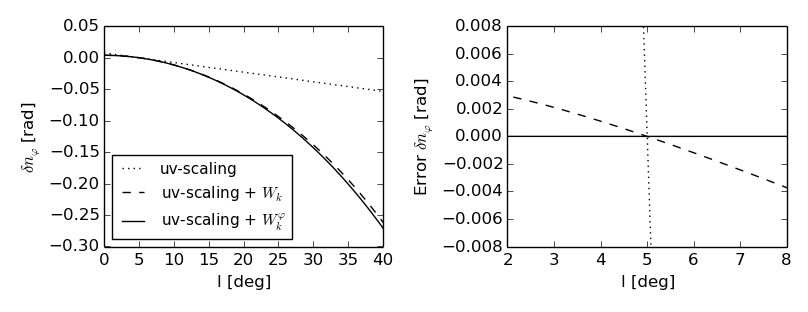}
\caption{\label{fig:Comparison} Comparison between the true
  $\delta  n_i$ term, and the first-order approximation (right), and
  residuals (left). The \citet{Kogan09} approximation breaks down away from the facet
  center (labeled as {$\u\v$-scaling}, dotted line). Applying
  classical W-projection together with a \citet{Kogan09} style
  coordinate transformation works better (dashed line), but a blind
  $3^{rd}-5^{th}$-order polynomial works best (solid line).
  \OMSOK{Not clear what the "true" value is, or is it effetively overlapping with thre solid line? Also,
  not clear what the polynomial is or does, I suppose you explain it later, but as a "naive" reader I'm confused.
  Also, which order is plotted, specifically? 3rd? 5th? Also, is the $y$ axis truly $\delta n$, or is it the full
  index of the exponent (i.e. times $w$)?}
  }
\end{center}
\end{figure}

As shown in Fig. \ref{fig:Comparison}, a more accurate approximation of the $\delta
n_\varphi=f\left\{\delta \l_\varphi,\delta \m_\varphi\right\}$ term may be obtained
by a fitting a low-order 2-dimensional polynomial


\def\iFacetCenter{\bm{\delta s_0}}
\def\BLs{\vec{b}_{pq}}

\begin{alignat}{2}
\delta \n_\varphi\approx&\left(\bm{\delta\l^k_\varphi}\right)^T\bm{\mathrm{P}_{k,\varphi}}\left(\bm{\delta\m^k_\varphi}\right)
\end{alignat}

\noindent where $\bm{\delta\l^k_\varphi}=\left[1,\delta\l_\varphi,\delta\l_\varphi^2,\hdots,\delta\l^k_\varphi\right]^T$ is a basis function vector for the $k^\mathrm{th}$-order 2-dimensional polynomial, and $\bm{\mathrm{P}_{k,\varphi}}$ is the matrix containing the polynomial coefficients. We can then write

\begin{alignat}{2}
\label{eq:FW0}
\BLs^T
\vec{\delta s}_\varphi&=
\u\delta \l_\varphi+\v\delta\m_\varphi+\w
\left(\bm{\delta\l^k_\varphi}\right)^T\bm{\mathrm{P}_{k,\varphi}}\left(\bm{\delta\m^k_\varphi}\right)\\
&\eqdefT{\overline{\mathrm{P}_{k}}}
\left(\u+\w \mathrm{P}_{k,\varphi}^{[10]}\right) \delta \l_\varphi+
\left(\v+\w \mathrm{P}_{k,\varphi}^{[01]}\right) \delta \m_\varphi\nonumber\\
\label{eq:FW1}
&\ \ \ \ \ \ 
+\w \left(\bm{\delta\l^k_\varphi}\right)^T\overline{\bm{\mathrm{P}_{k,\varphi}}}\left(\bm{\delta\m^k_\varphi}\right)\\
&\longeqdefT{\left(u',v'\right)}
u' \delta \l_\varphi+
v' \delta \m_\varphi
+\w \left(\bm{\delta\l^k_\varphi}\right)^T\overline{\bm{\mathrm{P}_{k,\varphi}}}\left(\bm{\delta\m^k_\varphi}\right)\\
&\longeqdefT{\MatFacetCoord}
\left(\MatFacetCoord\BLs\right)^T\vec{\delta s}_\varphi
+\w
\left(\bm{\delta\l^k_\varphi}\right)^T\overline{\bm{\mathrm{P}_{k,\varphi}}}\left(\bm{\delta\m^k_\varphi}\right)
\label{eq:CoordUVTranf}
\end{alignat}

\noindent where $\overline{\bm{\mathrm{P}_{k,\varphi}}}$ is equal to
$\bm{\mathrm{P}_{k,\varphi}}$ with the coefficients
$\mathrm{P}_{k}^{[01]}$ and  $\mathrm{P}_{k}^{[10]}$ in cells $[0,1]$
and $[1,0]$ zeroed. Based on this polynomial fit, we compute a series of convolutional 
kernels which we term \emph{\FW/-kernels} (for ``fitted $w$ kernels''), and apply them 
by exact analogy with W-projection. As in \citet{Kogan09}, we can see here
that the first order coefficient of the polynomial fit
$\bm{\mathrm{P}_{k,\varphi}}$ is equivalent to a $\w$-dependent
$(\u,\v)$ coordinate scaling. This has the effect of taking off the
main component of the $\w$-related phase gradient, and thereby
reducing the \FW/-kernels' support size (step from Eq. \ref{eq:FW0} to
Eq. \ref{eq:FW1}) that depends on (i) the $\w$-coordinate, (ii) the facet diameter and (iii) its location. In practice, if the facets are small enough (as it is the case when applying DDEs - see
Sec. \ref{sec:DDEFacetting}), the support of the \FW/-kernels is only marginally
larger than the spheroidal-only kernel.
The \FW/-kernels are computed per facet, per a given $w$-plane, as 

\begin{alignat}{2}
\mathcal{W}_{\mathcal{F}}\left\{\delta \l_\varphi,\delta \m_\varphi\right\}=&
\exp{
\left(-2 i\pi \frac{\nu}{c}\w
\left(\bm{\delta\l^k_\varphi}\right)^T\overline{\bm{\mathrm{P}_{k,\varphi}}}\left(\bm{\delta\m^k_\varphi}\right)
\right)}
\end{alignat}

\noindent In practice, a $3^\mathrm{rd}$ to $5^\mathrm{th}$-order polynomial is sufficient to accurately represent the 
$\w$-related phase variation across a given facet.



\newcommand{\MyWidth}{.8\columnwidth}
\begin{table}[t]\footnotesize
  \caption{\label{tab:symbols} Overview of the mathematical notations used throughout this paper}
\begin{tabular}{l@{~~}p{\MyWidth}}
\midrule
$\npix$\dotfill& 
number of pixels in the image domain\\
$\nvis$\dotfill& 
number of visibilities (per single frequency)\\
$\npar$\dotfill& 
number of spatio-spectral parameters to model the sky\\
$\Unity$\dotfill& 
The unit matrix\\
$\left<w_i f_i \right>_{\Omega} $\dotfill& The mean weighted sum of $f_i$ over a set of indices $\Omega$.
Weights are assumed to be normalized throughout.\\
$\vecb$\dotfill&
Index representing a baseline-time, i.e. a pair of antennas $pq$ at time $t$.\\
$\vecbnu$\dotfill&
Index representing a baseline-time-frequency (BTF) point.\\
$\Omega_\nu$ & The set of all BTFs $\BC$ (i.e., $\nvis$ elements) for a given frequency $\nu$. \\
$\VisBL$\dotfill& The 4-visibilities vector corresponding to $\BC$. \\
$\vecxnu$\dotfill&
The true sky vector (at frequency $\nu$) of size $\left(4\npix\right)$.\\
$\DirtyVecChan$\dotfill&
The dirty image vector (at frequency $\nu$) of size $\left(4\npix\right)$.\\
$\Fourier$\dotfill& 
The Fourier transform matrix of size $\left(\nvis \times \npix \right)$. Each column is the Fourier kernel mapping a given sky
coordinate to all visibilities.\\
$\SamplingBL$\dotfill& The matrix of size $\left(4\times 4n_v\right)$
representing the operator that selects 4 out of $4n_v$ visibilities.\\
$\MuellerImageDirBL{\vecs}$\dotfill& The $4\times 4$ Mueller-like matrix of a given
BTF point $\BC$, for direction $\vecs$.\\
$\MuellerImageBL$\dotfill& 
A block diagonal matrix of size $\left(4\npix\times 4\npix\right)$,
representing the DDE effect over the image domain for BTF point $\BC$. Each diagonal block $i$ is
a $4\times 4$ $\MuellerImageDirBL{\vecs_i}$ matrix.\\
$\TransferFunc$\dotfill& The direction-dependent transfer function at frequency
$\nu$, mapping the sky vector $\vecxnu$ onto the dirty image vector $\vecynu$\\
$\MuellerNorm$\dotfill& The image plane normalisation used to tranform
$\TransferFunc$ to a local convolution matrix.\\
$\widetilde{(.)}$\dotfill& An object affected by the image plane
normalisation $\MuellerNorm$.\\
$\WeightNu$\dotfill& Diagonal matrix of size $\left(4n_v \times 4n_v\right)$ representing the data
weighting for frequency $\nu$.\\
$\PSFMatBL$\dotfill& The PSF corresponding to a single BTF $\BC$. \\
$\TransferFuncPhiNoNorm$\dotfill& A transfer function mapping the
spatio-spectral sky model vector to the spectral dirty cube. \\
$\PiLocalChan$\dotfill& A $(\npix\times\npar)$ matrix describing the
transfer function between the
spatio-spectral sky model vector around pixel $i$ to the spectral
dirty cube at frequency $\nu$. \\
\bottomrule
\end{tabular}
\end{table}

\section{Imaging in the presence of Direction Dependent Effects}
\label{sec:DDEFacetting}

In this section, we describe, in terms of linear algebra, how the DDEs
are taken into account in the forward (gridding) and backward (gridding) imaging 
steps. 

Specifically, we describe how the dirty images
and associated PSFs are constructed from the set of
direction-time-frequency dependent Jones matrices. We show in
\Sec/ \ref{sec:DDEFacetting_FORWBACK} that,
in general, in the presence of baseline-time-frequency dependent
effects, the linear mapping $\TransferFunc$
(Eq. \ref{eq:TransferFunc} below) between the sky and the
dirty image is not a convolution operator. However, in
\Sec/ \ref{sec:DDEFacetting_IMCORR}, we describe a first-order image
correction that modifies this function into a direction-dependent convolution 
operator, under the condition that the Mueller matrices are
approximately baseline-time-frequency independent. As shown in
\Sec/ \ref{sec:DDEFacetting_DDPSF}, this correction is not sufficient, and the effective
PSF retains a directional variation. ``Local'' PSFs then have to be estimated per
facet. In this way, the normalized linear mapping
$\widetilde{\TransferFunc}$ (Eq. \ref{eq:ImageCorr} below) can be understood as a local convolution
operator.



\subsection{Forward and backward mappings}
\label{sec:DDEFacetting_FORWBACK}

In order to study the properties of the linear function, it is convenient to describe
this mapping from image to visibility space and back performed by the algorithm using linear algebra. 
For a given sample of 4 visibility products, Eq. \ref{eq:ME} can be
written in terms of a series of linear transformations:

\def\GlobalAChanBL{\bm{\mathcal{A}}_\BC}
\def\GlobalAChan{\bm{\mathcal{A}}_\nu}

\begin{equation}
\label{eq:ME_LinAlg_Block}
\begin{array}
{lcl}
\VisBL &=&
\SamplingBL \Fourier \MuellerImageBL \vecxnu
\eqdefT{\GlobalAChanBL}
\GlobalAChanBL\ \vec{x_{\nu}}
\end{array}
\end{equation}

\noindent where $\VisBL$ is the visibility 4-vector sampled by baseline $\vecb=(p,q,t)$ at frequency $\nu$ (for most of this
section, we assume a narrow-band scenario). If $\npix$ is the number of pixels in the sky model, the {\it true} sky image vector
$\vecxnu$ has a size of $4\npix$, and for each sky pixel
$i=1\dots\npix$, the four corresponding correlations\footnote{In practice the gridder and degridder of
  \DDFacet/ works on Stokes quantities that can be constructed from measured visibilities.} 
$(\textrm{XX},\textrm{XY},\textrm{YX},\textrm{YY})_i$ (or
$(\textrm{RR},\textrm{RL},\textrm{LR},\textrm{LL})_i$) are packed into $\vecxnu$ starting from index $4i$. 
Then, $\MuellerImageBL$ represents the DDEs, and is a $(4\npix)\times(4\npix)$ block diagonal
matrix. For any given image pixel $i$, the corresponding $4\times4$ block of 
$\MuellerImageBL$ is the Mueller-like\footnote{A Mueller matrix proper operates on Stokes vectors rather than 
visibility vectors. The Mueller matrix in this case would be given by $\mathbf{S}^{-1}\MuellerImageDirBL{\vecs_i}\mathbf{S}$, where
$\mathbf{S}$ is the $4\times4$ conversion matrix mapping Stokes vectors to visibility vectors. See for example \citet{Hamaker96,Oleg11}}
matrix associated with the pixel direction $\vecs_i$:
$\MuellerImageDirBL{\vecs_i}=\JonesMat^*_{q}(t,\nu,\vec{s}_i)\otimes\JonesMat_p(t,\nu,\vec{s}_i)$. 
$\Fourier$ is the Fourier transform operator of size
$(4\nvis)\times(4\npix)$. Each of its $(4\times4)$ blocks is a
scalar matrix, the scalar being the kernel of the Fourier basis
$k^{\vecs_i}_{(pq),t\nu}$ (see Eq. \ref{eq:kterm}). The matrix
$\SamplingBL$ is the sampling matrix, size $4\times (4\nvis)$, which selects the 4 visibilities corresponding to $\BC$

For the full set of $\nvis$ 4-visibilities associated with channel $\nu$, which we designate as $\Omega_\nu$, 
($\BC\in\Omega_\nu$ then means that the $\BC$ index can be taken to represent a visibility index from $1$ to $\nvis$), 
we can stack $\nvis$ instances of Eq. \ref{eq:ME_LinAlg_Block} to write the forward (image-to-visibility) mapping as:

\begin{equation}
\label{eq:ME_LinAlg}
\begin{array}
{lcl}
\VisChan &=&
\begin{bmatrix} 
\vdots \\ 
\GlobalAChanBL\\ 
\vdots \\ 
\end{bmatrix}
\vecxnu
\eqdefT{\GlobalAChan}
\GlobalAChan\vecxnu
\\
\end{array}
\end{equation}

Note that $\GlobalAChan$ represents the ``ideal'' mapping from images to visibilities, in the sense that a 
unique DDE is applied at every pixel (ignoring the approximation inherent to pixelizing the sky, we can say
that $\GlobalAChan$ represents the true instrumental response). Implementing $\GlobalAChan$ directly in the forward 
(modeling) step of an imager would be computationally prohibitive: it is essentially a DFT (Direct Fourier Transform) with pixel-by-pixel 
application of DDEs. Existing approaches therefore construct some FFT-based (Fast Fourier Transform) approximation to $\GlobalAChan$. The 
convolutional function approach, i.e. AW-projection, approximates $\GlobalAChan$ by a single FFT followed by convolutions 
in the $uv$-plane during degridding. The facet-based approach of the present work segments the sky $\vecxnu$ into 
facets, then does an FFT per-facet, while applying a constant DDE $\MuellerImageDirBL{\vecs_\varphi}$ (where $\vecs_{\varphi}$ 
is the direction of facet $\varphi$). The resulting approximate forward operator, $\widehat{\GlobalAChan}$, 
becomes exactly equal to $\GlobalAChan$ in the limit of single-pixel facets (see \Sec/~\ref{sec:perfacet} for 
a further discussion).

\subsection{Forming the dirty image}

Since $\GlobalAChan$ is generally noninvertible, imaging algorithms tend to construct the adjoint operator 
$\GlobalAChan^H$, or some approximation thereof $\widehat{\GlobalAChan^H}$, to go back from the visibility domain to the image domain\footnote{This
can be motivated as follows: for any given matrix $\bm{A}$, the null spaces $\KER{\bm{A}}$ and $\KER{\bm{A}^H \bm{A}}$ are identical. Therefore, applying the adjoint operator $\bm{\mathcal{A}}^H$ to go back to images from visibilities 
preserves all information not destroyed by $\bm{\mathcal{A}}$.}. This amounts to forming the so-called dirty image.

In the framework of facets and DDE calibration, we obtain what is at best an estimate $\widehat{\GlobalAChan}$ (due to 
finite facet sizes, and also calibration errors), and therefore the adjoint operator being applied is also an approximation. 
The same applies to convolutional gridding approaches. For the purposes of this section, however, let us assume that the 
approximation is perfect.  We then have the following for the dirty image vector $\vecynu$:

\begin{alignat}{2}
\label{eq:TransferFunc}
\DirtyVecChan &=\GlobalAChan^H \WeightNu \VisChan\\
\label{eq:TransferFunc1}
 &=\GlobalAChan^H \bm{\mathcal{W}}_\nu
\GlobalAChan\ \vecxnu\ \ 
\eqdefT{\bm{\mathcal{T}}} \bm{\mathcal{T}}\ \vecxnu
\\
&=\left<
w_{\BC}
\MuellerImageBLH
\Fourier^H \SamplingBL^H \SamplingBL \Fourier 
\MuellerImageBL
\right>_{\OmegaB}\ \vecxnu\\
&\eqdefT{\PSFMatBL}\left<
w_{\BC}
\MuellerImageBLH
\PSFMatBL
\MuellerImageBL
\right>_{\OmegaB}\ \vecxnu
\label{eq:TransferFunc2}
\end{alignat}

\noindent where 
$\WeightNu$ is a diagonal matrix containing the set of weights $w_{\BC}$ at frequency $\nu$. Note that the 
weighted sum comes about due to the block-column of Eq.~\ref{eq:ME_LinAlg} being left-multiplied by its conjugate, a block-row.

For each $\BC$, the matrix $\PSFMatBL=\Fourier^H \SamplingBL^H \SamplingBL \Fourier$ is a convolution, as a direct 
consequence of the Fourier convolution theorem. This matrix represents the convolution of the sky by the PSF 
corresponding to a single $uv$-point (i.e. a single fringe). In the absence of DDEs 
($\MuellerImageBL\equiv\Unity$), the linear mapping $\TransferFunc$ can be written as a a weighted sum of such
matrices, and is therefore also a convolution:

\begin{equation}
\vecynu = \left< w_{\BC} \PSFMatBL \right>_{\OmegaB} \vecxnu 
\eqdefT{\bm{\mathcal{C}}} 
\bm{\mathcal{C}} \vecxnu
\label{eq:psf}
\end{equation}

This is just the familiar result that in the absence of DDEs, the dirty image is a convolution of the apparent sky by a PSF.

Below, we show that in the presence of DDEs (even corrected-for DDEs), this relationship generally ceases to be a true convolution. 
We will also show that, under certain conditions, the mapping can be modified (at least approximately) into a 
\emph{local convolution}, i.e. one where the PSF varies only slowly with direction. This distinction is important: most minor-loop
deconvolution algorithms such as CLEAN either assume the problem is a true (global) convolution, or can be trivially 
modified (at least in the faceted approach) to deal with a local convolution problem, i.e. a position-dependent (per facet) PSF.

\subsection{Toeplitz matrices and convolution}

Any matrix $\mathbf{C}$ representing a one-dimensional convolution is Toeplitz, and vice versa. A Toeplitz matrix is a 
matrix in which each descending diagonal is constant, i.e. $C_{ij}=C_{i+1,j+1} \equiv c_{i-j}.$ We now show 
that a similar property, which we'll call \emph{Toeplitzian}, can be defined for convolution of 2D, 
4-polarization images. We can then discuss how DDEs break the convolution relationship by making the 
matrix representing the tranfer function less Toeplitzian.

First, consider matrices that represent 2D convolution of scalar (unpolarized) images. The pixel ordering, i.e. 
the order in which we stack the pixels of a 2D image into the image vectors $\vec{x}$ and $\vec{y}$, induces a 
mapping from vector index $i$ to pixel coordinates $(l_i,m_i)$. Given a fixed pixel ordering, consider a 
matrix $\mathbf{C}$ whose its elements are constant with respect to a translation of pixel 
coordinates, i.e. $C_{ij} = C_{i^\prime j^\prime}$ for all $ij$ and $i^\prime j^\prime$ such that 
$l_i-l_j=l_{i^\prime}-l_{j^\prime}$ and $m_i-m_j=m_{i^\prime}-m_{j^\prime}$. There then exists
a function of pixel coodinates $c(l,m)$ such that for all $ij$

\begin{equation}
\label{eq:2dtoeplitz}
C_{ij} = c(l_i-l_j,m_i-m_j) = c(\Delta l_{ij},\Delta m_{ij}),
\end{equation}

\noindent and it is then easy to see that applying the $\bm{C}$ matrix to the image vector $\vec{x}$ corresponds to a 
2D convolution of the corresponding image by $c$, and vice versa. For an $n\times n$ image, assuming the conventional 
pixel ordering of stacked columns (or rows), the matrix $\bm{C}$ is composed of $n\times n$ blocks, each block 
being an $n\times n$ Toeplitz matrix. Each constant descending diagonal in each such block represents a
constant pixel separation $\Delta l, \Delta m$. In other words, $C_{ij}$ is constant for any pair of pixels 
having the same pixel separation $\Delta l, \Delta m$.

To generalize this to 4-polarization images, we simply replace $C_{ij}$ in Eq.~\ref{eq:2dtoeplitz} by 
a $4\times4$ scalar matrix. Our general \emph{Toeplitzian} matrix is then composed of $n\times n$ blocks, each block 
being a $4n\times 4n$ Toeplitz matrix composed of $4\times4$ scalar matrices. Each column of such matrix represents 
the convolution kernel (or PSF), shifted to the position of the appropriate image pixel.

The linear function defined by the PSF $\PSFMatBL$ or $\bm{\mathcal{C}}$ is Toeplitzian, with 1 on the main diagonal
(corresponding to the peak of the PSF). We focus on two regimes in which a matrix becomes non-Toeplitzian. 
The first one is simple, when $C_{ij}$ in Eq.~\ref{eq:2dtoeplitz} is constant to within a $4\times 4$ 
per-column scaling factor $M_j$. This correponds to an attenuation of the image by $\MuellerImage$, followed by a convolution:

\begin{equation}
\vecynu = \bm{\mathcal{C}} \MuellerImage \vecxnu \eqdefT{\widetilde{\vecxnu}} \bm{\mathcal{C}} \widetilde{\vecxnu}
\end{equation}

This regime arises when trivial (i.e. non time-baseline dependent) DDEs are present and not accounted for when 
forming the dirty image. $\MuellerImage$ can be factored out of the sum in Eq.~\ref{eq:TransferFunc2} and 
absorbed into the \emph{apparent sky} $\widetilde{\vecxnu}$. In this case we can still talk of the PSF 
shape being constant across the image.

The more complex regime arises when the mapping is non-Toeplitizian in the sense that the shape 
of the PSF changes across the image. This naturally arises when nontrival DDEs are present and not accounted for, and the dirty image is the weighted sum of the sky affected by baseline-dependent DDE

\begin{equation}
\vecynu = \left< w_{\BC} \PSFMatBL \MuellerImageBL \right>_{\OmegaB} \vecxnu 
\end{equation}

More subtly, even if DDEs {\bf are} perfectly known and accounted for in $\bm{\mathcal{A}}^H$, the resulting 
function is, generally, not a convolution, in the sense that the shape of the PSF becomes 
direction-depedent. This is obvious in the case of nonunitary $\MuellerImageBL$
(since its amplitude essentially appears twice in Eq.~\ref{eq:TransferFunc2}, and the resulting dirty image 
requires renormalization -- we will return to this again below). Less obvious is that this holds, generally,
even for unitary $\MuellerImageBL$. Consider the simple case of a scalar, unitary DDE (i.e. a phase term affecting
both polarizations equally). This corresponds to a diagonal $\MuellerImageBL$ 
with $M_i = \mathrm{e}^{\imath \psi_i}$ on the diagonal. If the matrix elements of $\PSFMatBL$ are given by $C_{ij}$,
then each element of $\MuellerImageBLH \PSFMatBL \MuellerImageBL$, i.e. the response at dirty image pixel
$j$ to a source at pixel $i$ (i.e. the PSF sidelobe response), is given by

\begin{equation}
\label{eq:Cij:wproj}
C^\prime_{ij} = M^*_i C_{ij} M_j = C_{ij} \mathrm{e}^{\imath{(\psi_j-\psi_i)}} 
\end{equation}

It is easy to see that the (Toeplitzian) condition of Eq.~\ref{eq:2dtoeplitz} is only satisfied if 
$\psi_j-\psi_i$ is constant for any pair of pixels having the same pixel separation $\Delta l, \Delta m$. This
condition is only true for a linear phase slope over the image.

We have shown that here all nontrivial DDEs, including unitary ones, with the exception of linear phase slopes, 
generally result in a direction-dependent PSF even when perfectly known and accounted for via 
Eq.~\ref{eq:TransferFunc}. Note that this equation (or some approximation thereof) is applied by all 
existing imagers. If we consider the $w$-term as a DDE \citep[see, e.g.,][]{Oleg11}, we can see that 
W-projection and W-stacking also represent approximations of Eq.~\ref{eq:TransferFunc}, and therefore still yield
a direction-dependent PSF.

\subsection{Loss of local convolution property and nonuniform noise reponse}
\label{sec:LossCovProp}

Equations~\ref{eq:2dtoeplitz} and \ref{eq:Cij:wproj} give us a framework in which we can reason about the degree of
direction-dependence in the PSF. The pixel separation $\Delta l, \Delta m$ 
corresponds to the PSF sidelobe at $C_{ij}$. Thus, the direction-dependence of a particular PSF sidelobe
$\Delta l,\Delta m$ is determined by the variation of the Mueller matrix across the image on a length 
scale of $\Delta l,\Delta m$. For direction-dependent effects that are locally approximately linear (i.e. close
to the form of Eq.~\ref{eq:Cij:wproj}), the problem is locally a convolution. As long as this is true, and assuming 
$\MuellerImageBL$ is known, one could in principle incorporate knowledge of a local direction-dependent PSF into 
the minor cycle deconvolution algorithm, using the linear function defined above to form up the dirty images. In the 
context of facet imaging this seems straightfoward, as we can simply compute a PSF per facet (see below). 
However, if the Mueller matrices are nonunitary, $\TransferFunc$ has two very undesirable properties.

Firstly, as is clear from Eq.~\ref{eq:Cij:wproj}, the PSF sidelobe response $C^\prime_{ij}$ is coupled to 
$\MuellerImageBL$ at both positions $i$ and $j$. \OMSOK{Particularly in the faceted case, this makes it 
impossible to use an approximate piecewise-constant PSF in deconvolution. Consider, for example, a 
piecewise-constant DDE that perfectly matches our faceting, and consider a source at pixel $j$ moving 
around one particular facet. For a fixed pixel separation $\Delta l,\Delta m$, the PSF sidelobe response at 
pixel $i$ will then depend on what facet $i$ falls into, and this can change as we move pixel $j$ around 
its ``home'' facet. In other words, the effective PSF of pixel $j$ is not even constant across a facet.}
\OMSOK{My comment to Cyril: I think this is what you're actually trying to get at when you talk about
bringing the mapping back to a direction-dependent local convolution. The PSF needs to be approximately 
constant across a facet, and that's what normalization gives you. In the limit of one pixel = one 
facet, you don't actually
need normalization, since it's not a convolution with or without normalization, and you'd be using a per-pixel 
PSF anyway!} Ideally, we would like to decouple the PSF sidelobe response from the DDE at position $i$.

Secondly, consider the thermal noise response in the dirty image given by $\TransferFunc$. Thermal 
noise can be assumed to be independent and identically distributed Gaussian in the visibilities $\vec{v}_{\BC}$. If $\vec{a}$ is a vector of random variables and 
$\vec{b}=\bm{B}\vec{a}$, then the covariance matrices of the two vectors are related by
$\Cov{\vec{b}}=\bm{B}\Cov{\vec{a}}\bm{B}^H$. Applying this to Eq.~\ref{eq:TransferFunc}, and using 
$\sigma^2_{\BC}$ for the variance of the real and imaginary parts of each visibility, we get

\begin{equation}
\label{eq:covy}
\Cov{\DirtyVecChan} = \left< w^2_{\BC} \sigma^2_{\BC} \MuellerImageBLH \PSFMatBL \MuellerImageBL \right >_{\OmegaB},
\end{equation}

\noindent i.e. the thermal noise in the dirty image is not spatially uniform. In particular, the variance 
(of the four polarization products) at each pixel $i$ is given by 
$\left< w^2_{\BC} \sigma^2_{\BC} \MuellerImageBLH_i \MuellerImageBL_i \right >_{\OmegaB}$.

\subsection{Image-plane corrections}
\label{sec:DDEFacetting_IMCORR}

\OMSOK{Up to here it's OK, but now I'm confused. If I can't even convince myself, how do you convince a referee?
You do it... I've left the following text unaltered, but here are my thoughts about it.
Yoy say baseline-time-frequency independent, but what does frequency have to do with it (we explicitly say we look at 
a narrowband case, per gridding band).
You postulate in 
that the transfer operator can be approximated by an image-plane effect (diagonal 
$M$), times the convolution operator, times the conjugate image-plane effect (diagonal $M^H$). In the case of 
a baseline-time independent $M$ this is exact. It's not clear to me where the approximation breaks down. We can ignore
this issue entirely and say: if such an diagonal approxmation exists, then it must be given by 
because it must be the square root of the diagonal of $T$. So we calculate the 
correction factor assuming the approximation exists (because {\bf if} it exists, {\bf then} that's what the correction
factor must look like), and hope for the best. Like I said, I find 
this logic Jesuitical, so I'll leave it to you to finesse it.
I don't like the term ``direction-dependent convolution matrix'' when it's just attenuation of the sky, 
followed by a true convolution.``DD convolution'' makes me think of a DD PSF shape.
Unitary vs. nonunitary Jones matrices: I think this is covered by my text above, so can probably be taken out.

Finally, and this one bugs me the most.
is not quite correct. If you look at the 
equation before it, the product contains two $w$'s and two inverse square roots of $w$. So 
should contain $w$ not $w^2$ surely? And if this equation is wrong, does this mean that the 
image-plane correction doesn't actually give us flat noise?  
What does the code use, anyway? $w$ or $w^2$?
End of problem list.}

We show in this section that when $\MuellerImageBL$
is approximately baseline-time independent (at a given frequency), we can find a dirty
image correction that brings $\bm{\mathcal{T}}$ back to a
direction-dependent convolution operator. This is a reasonable assumption in general
if the fractional bandwith of the data chunk is small
enough\footnote{In \DDFacet/ the data is internally imaged into
  frequency-chunked spectral cubes, and the correction described here
  (Eq. \ref{eq:ImageCorr}) is done on a user-defined but ideally small
  bandwidth.}, and in this case we can write

\begin{alignat}{2}
\label{eq:TMTilde}
\bm{\mathcal{T}}\simeqdefT[3]{\MuellerNorm}& 
\MuellerNorm^H
\left<
w_b\PSFMatBL
\right>_{\OmegaB}
\MuellerNorm\\
&\eqdefT{\bm{\mathcal{C}_S}}
\MuellerNorm^H
\bm{\mathcal{C}_S}
\MuellerNorm
\end{alignat}


\noindent We can see from Eq. \ref{eq:TMTilde} that we can construct a
modified normalized image

\begin{alignat}{2}
\label{eq:ImageCorr}
\DirtyVecNormChan =&\left(\MuellerNorm^{H}\right)^{-1}\TransferFunc \vec{x_{\nu}} 
\eqdefT{\TransferFuncNorm}\widetilde{\TransferFunc}\vec{x_{\nu}}
\end{alignat}

\noindent where $\widetilde{\TransferFunc}$ is the normalized mapping. The columns $i$ of $\widetilde{\TransferFunc}$ are then
the PSF of a source centered on pixel $i$, and they only differ one to
the other by a matrix product. In other words, the PSF centered on pixel $i$ is
the same as the PSF centered on pixel $j$ to within a
constant. Strictly speaking $\widetilde{\TransferFunc}$ is not a convolution matrix,
but we will refer to it as a direction-dependent convolution matrix. An
alternative way to look at this is to write
$\DirtyVecNormChan=\bm{\mathcal{C}_S}\widetilde{\vec{x_{\nu}}}$ where
$\widetilde{\vec{x_{\nu}}}=\MuellerNorm\vec{x_{\nu}}$ is the apparent
beam-attenuated sky.

In order to obtain $\MuellerNorm$, we can see from Eq. \ref{eq:TMTilde} that
although $\TransferFunc$ is not block diagonal, each $i$-th
$(4\times4)$ block on its diagonal is 

\begin{alignat}{2}
\TransferFunc[i,i]=&\left<w_b \smallM^H_{b,i}\smallM_{b,i}\right>_{\OmegaB}
\end{alignat}

Assuming approximate baseline-time independence at $\nu$ of the
direction-dependent local convolution function (i.e.\ 
$\TransferFunc[i,i]\approx\widetilde{\smallM_i}^H\widetilde{\smallM_i}$)),
we get

\begin{alignat}{2}
\label{eq:MuellerNorm}
\MuellerNorm\widehat{=}&\sqrt{\left<w_b \MuellerImageBL^H\MuellerImageBL\right>_{\OmegaB}}
\end{alignat}

\begin{figure*}[ht!]
\begin{center}
\includegraphics[width=5.5cm]{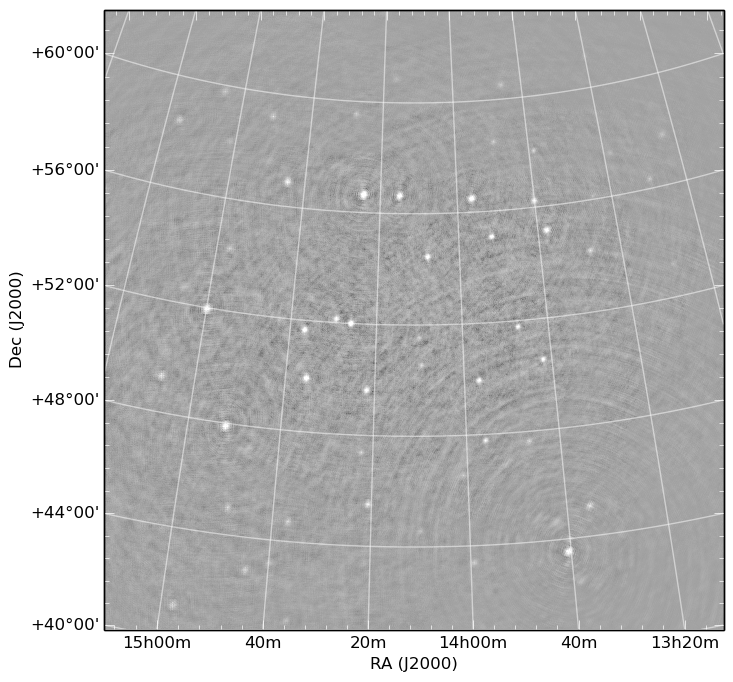}
\includegraphics[width=5.5cm]{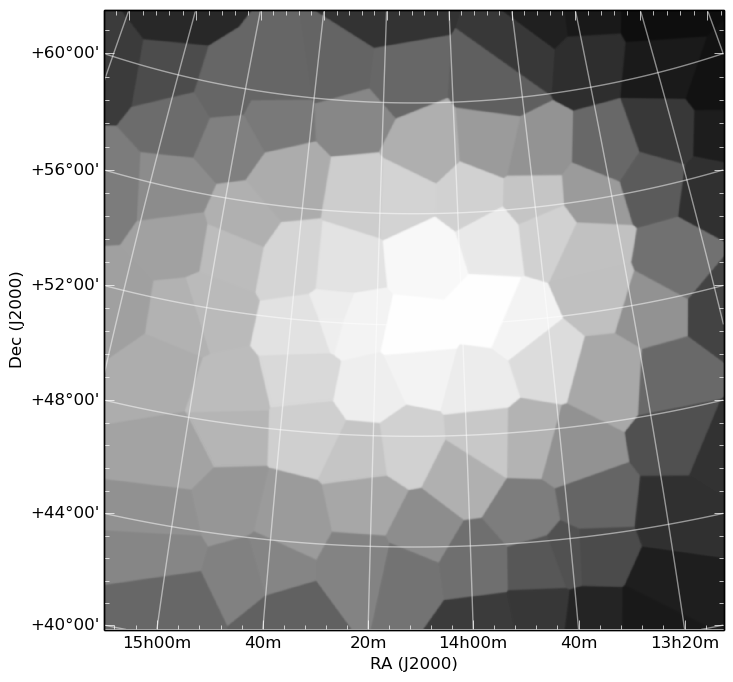}
\includegraphics[width=5.5cm]{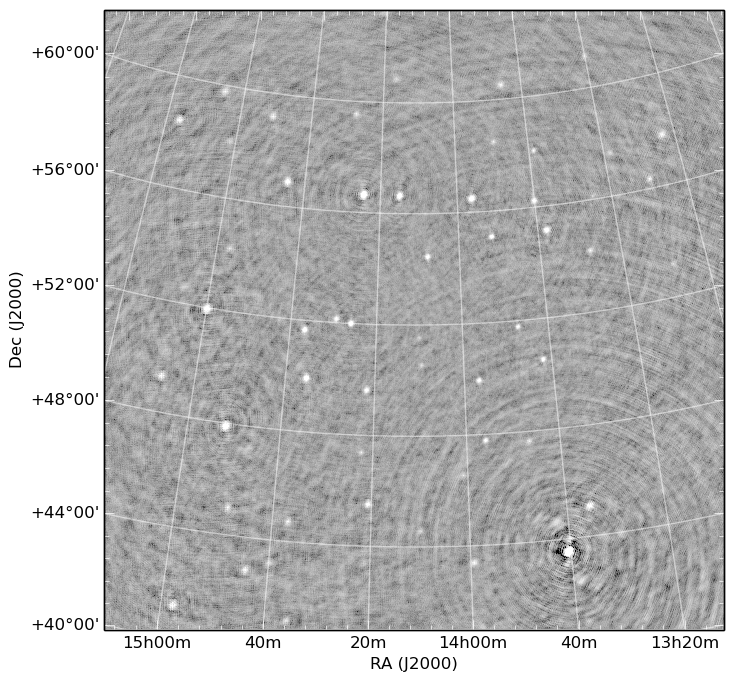}
\end{center}
\caption{\label{fig:ImageCorr} The nonunitary Mueller matrices in
  $\mathcal{A}$ cause the mapping in
  Eq. \ref{eq:TransferFunc1} to not be a convolution operator. The
  left panels shows the dirty image $\DirtyVecChan$ obtained after
  applying $\mathcal{A}^H$ to the visibilities. The image plane
  correction $\MuellerNorm$ is displayed in the central
  panel and the corrected image $\DirtyVecNormChan$ is shown in the right
  panel. As explained in \Sec/ \ref{sec:DDEFacetting_IMCORR} the modified mapping is
  approximately a
  direction-dependent local convolution function.}
\end{figure*}

If the assumption in Eq. \ref{eq:TMTilde} holds (definition of $\MuellerNorm$), then the image plane correction exists, and it is given by 
Eq. \ref{eq:MuellerNorm}. Furthermore we take into account the
deviation from this approximation by using the local PSF
(Sec. \ref{sec:DDEFacetting_DDPSF}) in our deconvolution
algorithms (Sec. \ref{sec:WBDeconv}). Applying the $\MuellerNorm$
correction in Eq. \ref{eq:covy}, the normalized
image-plane pixel covariance $\Cov{\DirtyVecNormChan}$ becomes



\begin{equation}
\label{eq:covyCorr}
\Cov{\DirtyVecNormChan} = \left< w^2_{\BC} \sigma^2_{\BC} \PSFMatBL \right >_{\OmegaB},
\end{equation}

\noindent and $\Cov{\DirtyVecNormChan}$ is spatially uniform.

In practice, the Mueller blocks in $\MuellerImageBL$ are assumed to be
diagonally dominant and are reduced to scalar matrices when computing
$\MuellerNorm$.




\subsection{Direction-dependent PSFs}
\label{sec:DDEFacetting_DDPSF}

As shown above, the combined effects of (i) baseline-time-frequency dependence of the
DDEs, and (ii) decorrelation cause the linear mappings
$\TransferFunc$ and $\TransferFuncNorm$ not to
be exact convolution matrices. Specifically, the large fractional bandwidth
makes the beam pattern vary significantly toward the edge of the field, and the
effective PSF is also direction-dependent. All modern imagers are indeed affected by problem (i) in the minor
cycle, and problem (ii) in both the major and minor cycles, and so will
produce morphologically biased results away from the pointing
center. In this section we describe how \DDFacet/ takes into account
and compensates for these effects.

\subsubsection{Effect of decorrelation}
\label{sec:DecorrelationDDPSF}

It follows from Eq. \ref{eq:ME} that any source in the sky corresponds
to a complex
vector rotating in the {\it uv}-domain and any visibility measurement
is an averaged value over that domain. This fact causes the
amplitude of the averaged vector to decrease (in the extreme case in which
the phase of the complex vector ranges over $[-\pi,\pi]$ in the domain
of averaging, the average vector amplitude can be zero). This effect
is known as decorrelation and is described in much detail by
\citet[][and references therein]{Smear99,tms,Oleg11,atemkeng2016}. One can see from
Eq. \ref{eq:ME} that the magnitude of decorrelation depends on (i)
the baseline coordinates and (ii) the distance of the source to the phase
center, causing the effective PSF to be direction-dependent. This effect is a
direct image-domain consequence of baseline and direction-dependent
decorrelation, and is known in the literature as {smearing}.

The effective mapping is therefore direction-dependent, and
no imaging and deconvolution can take this effect into account. This has the
direct effect of incorrectly estimating the source's morphology, and
the error gets worse as the source gets further away from the phase
center. Since the longest baselines are most affected, decorrelation is
minimized by accepting a small decorrelation (e.g. a few per cent
decrease in the ratio to peak to integrated flux density) at the edge of the
field.

A major strength of a facet-based imaging and deconvolution framework is
that we can take decorrelation into account in quite an easy way by
computing a PSF per facet. While computing the PSF, each unit visibility is multiplied by the
factor $\gamma_{pq,t}$, defined as

\begin{alignat}{2}
\gamma_{pq,t\nu} =&\frac{\sin{\left(\phi_{pq,t}\right)}}{\phi_{pq,t}}\frac{\sin{\left(\psi_{pq,t}\right)}}{\psi_{pq,t}}\\
\textrm{with } \phi_{pq,t\nu}=&\pi \frac{\Delta_{\nu}}{c}\vec{b}_{pq,t}^T\left(\vec{s}-\vec{s}_0\right)\\
\textrm{and } \psi_{pq,t\nu}=&\pi
\frac{\nu}{c}\ddt{\vec{b}_{pq,t}}^T\left(\vec{s}-\vec{s}_0\right)\Delta_t
\end{alignat}

\noindent where $\Delta_t$ and $\Delta_{\nu}$ are the time and
frequency size for the domain over which the given visibility has been
averaged, and $\ddt{\vec{b}_{pq,t}}$ is the {\it speed} of the baseline
in the uv domain. Conversely, in the forward step of major cycle, $\gamma_{pq,t\nu}$ can be applied to the model visibilities on a per-facet basis. This allows decorrelation to be properly accounted for both in the minor and major cycles.

\subsubsection{Per-facet PSFs}

In the facet approach, it is staightforward to compute a per-facet PSF that takes all of the above effects into account
during deconvolution. We compute a PSF per facet $\phi$ for a point source following

\begin{alignat}{2}
\DirtyVecNormChan_{\vec{1}} =&\widetilde{\TransferFunc^{\phi}}\vec{0_1}
\end{alignat}

\noindent where $\widetilde{\TransferFunc^{\phi}}$ is the local
convolution function
function in facet $\phi$, and $\vec{0_1}$ is a vector containing zeros everywhere except 
the central pixel, which is set to the value $\{I,Q,U,V\}=\{1,0,0,0\}$ Jy. 
Fig.~\ref{fig:PSF} shows the PSF evaluated for a source in two different facets.

\def\FigWidth{3.5}
\begin{figure}[h!]
\begin{center}
\includegraphics[width=\FigWidth cm]{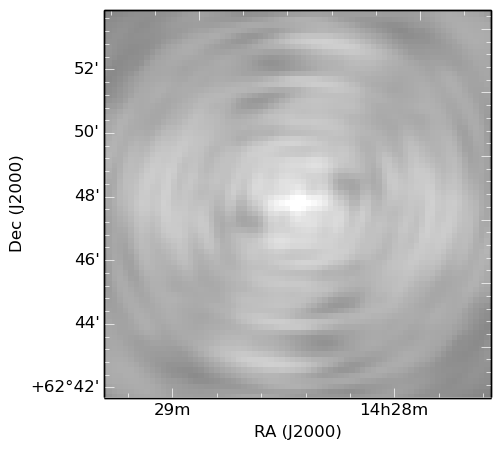}
\includegraphics[width=\FigWidth cm]{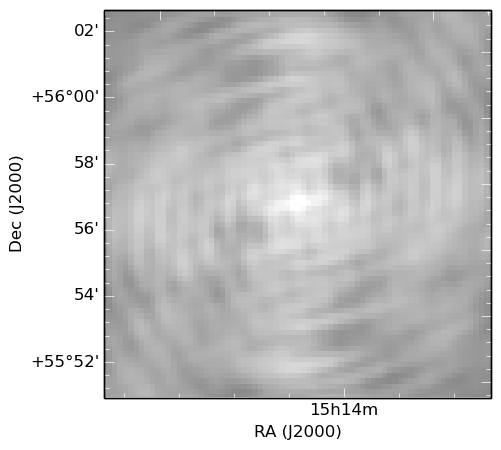}
\end{center}
\caption{\label{fig:PSF} This figure shows the PSF estimated at
  various locations of the image plane even after the transformation
  described in Eq. \ref{eq:ImageCorr} is applied. The net local
  convolution function significantly varies,
  and this effect is taken into account by computing a PSF per facet.}
\end{figure}

Note that in the full-polarization case, i.e. given DDEs with a nontrivial polarization response (nondiagonal, 
or at least nonscalar Mueller matrices), it is in principle incorrect to speak of one PSF. All four 
Stokes components are, in general, convolved with different PSFs, and there are also ``leakage PSFs'' that transfer
power between components. A fully accurate description of the local convolution relationship therefore requires 
that 16 independent PSFs be computed, with all the consequent expense (i.e. 16 separate gridding operations for the 
PSF computation). In practice, we limit ourselves to computing the Stokes $I$ PSF, and, during the minor 
cycle of deconvolutin, assume that the other Stokes component PSFs are the same, and treat leakages as 
negligible, trusting in the major cycle to correct the effect. The impact of this approximation 
on polarization deconvolution is a topic for future study.

\subsection{In-facet errors}
\label{sec:perfacet}

As explained in \Sec/ \ref{sec:Intro}, the facet-based imaging and
deconvolution framework presented here has the disadvantage of taking
DDEs into account in a discontinuous manner in the image
domain. Indeed, within a direction-dependent facet, DDEs are assumed to
be constant while they continuously vary. This is typically the case
for beam effects that vary very quickly, especially around the half power
point. We show in this section that this effect can be partially
accounted for by applying a spatially smooth term to the image $\DirtyVecNormChan$.



In this section we estimate the flux density error across a given
facet $\varphi$ that arises due to the fact that the Jones matrix has been
assumed to be spatially constant. Following
Eq. \ref{eq:ME_LinAlg_Block}, the residual visibility on a given
baseline $\vec{b}$ can be written as

\begin{equation}
\label{eq:ResidualVis}
\begin{array}
{lcl}
\ResVisBL &=&
\SamplingBL \Fourier 
\left(
\MuellerImageBL \vec{x_{\nu}}
-
\MuellerImageBLConst \EstimatedSky
\right)
\end{array}
\end{equation}

\noindent where $\MuellerImageBLConst$ is a $(4\npix)\times(4\npix)$ block diagonal
matrix which represents the direction-independent Jones matrix that
has been assumed for that facet, and $\EstimatedSky$ is the sky that
has been estimated. We assume the deconvolution algorithm is subject to
an $\mathcal{L}_2$-norm constraint, and


\begin{alignat}{2}
\chi^2 =& \displaystyle \sum_{\vec{b}}\left(
\MuellerImageBL \vec{x_{\nu}}-\MuellerImageBLConst \EstimatedSky
\right)^H
\PSFMatBL
\left(
\MuellerImageBL \vec{x_{\nu}}-\MuellerImageBLConst \EstimatedSky
\right)
\end{alignat}

\noindent is minimized, giving

\begin{alignat}{1}
\frac{\partial \chi^2}{\partial \EstimatedSky} =\vec{0}=\displaystyle\sum_{\vec{b}}& 
2\MuellerImageBLConstT \PSFMatBL\MuellerImageBLConst  \EstimatedSky \nonumber\\
&-2\MuellerImageBLConstT \PSFMatBL\MuellerImageBL\vec{x_{\nu}}
\end{alignat}

\noindent and therefore

\begin{alignat}{1}
\nonumber
\vec{x_{\nu}}=&
\left(\displaystyle\sum_{\vec{b}}\MuellerImageBLConstT
\PSFMatBL\MuellerImageBL\right)^{-1}\\
&\ \ \ \left(\displaystyle\sum_{\vec{b}}\MuellerImageBLConstT
\PSFMatBL\MuellerImageBLConst\right)
\EstimatedSky
\end{alignat}

As in \Sec/ \ref{sec:DDEFacetting_IMCORR}, assuming the
$\MuellerImageBL$ and $\MuellerImageBLConst$ are
baseline-time independent at $\nu$ we get

\begin{alignat}{1}
\vec{x_{\nu}}=&
\MuellerImage^{-1}
\MuellerImageConst
\EstimatedSky
\end{alignat}

One can see that when no DDEs are being applied during deconvolution
($\MuellerImageConst=\textrm{\bf{I}}$, as is traditionally done in radio
astronomy), one can correct the fluxes by applying a {\it smooth beam}
correction in the image domain.

\section{Wideband deconvolution}
\label{sec:WBDeconv}

In this section, we describe how we solve for the sky in the local
deconvolution problem as well as the global inverse
problem\footnote{In an abuse of language we can also call inverting
  $\widetilde{\TransferFunc}$ a deconvolution problem, although strictly
  speaking it is not a convolution operator.}
$\DirtyVecNorm=\widetilde{\TransferFunc}\vec{x}$. We present two multiscale wideband
deconvolution algorithms that iteratively estimate the underlying
{true} sky. In contrast to the calibration problem, the
deconvolution problem is linear, but is strongly ill-conditioned. A
wide variety of algorithms have been developed to tackle the
conditioning issue.

The first and largest family of deconvolution algorithms in radio
interferometry is based on compressive sampling theory (or compressive
sensing), and assumes the sky can be fully described by a small number
of coefficients in a given dictionary\footnote{In contrast to a
  basis, the decomposition of a vector in these dictionaries is not
  necessarily unique.} of functions (a {\it sparse representation}). The
dictionary of functions can be, but is not necessarily, a basis function from deltas to
shapelets. In practice and for a given dataset, a
specific convex solver is used to estimate
the coeffiscient associated to the functions of the dictionary. The
cost function is often an $\mathcal{L}_1$-norm subject to an
$\mathcal{L}_2$ constraint.
The widely used {\sc clean} algorithm is one of
those\footnote{Although {\sc clean} was written before compressive
  sampling theory had been described.} solvers, but we can also mention {\sc
  moresane} \citep{Dabbech15}, or {\sc sasir} \citep{Garsden15}. Each one of these methods 
  uses a specific
solver to estimate the coefficients associated with a given
dictionary. The second family of algorithms deals
with ill-conditioning using Bayesian inference.

Only a few existing algorithms are able to accurately estimate flux
densities as well as intrinsic spectral properties (while taking Jones
matrices into account). The most efficient and widely used of these is the
\MTMSCLEAN/ algorithm \citep[for multi-term multi-scale, see][and references
  therein]{Rau11}. \citet{Bhatnagar13} have extended this algorithm in
order to take time-frequency dependent DDEs into account. The drawbacks
of this algorithm combination are that (i) since each Taylor
coefficient image stacks information from potentially large
fractional bandwidth, $\widetilde{\TransferFunc}$
(Eq. \ref{eq:ImageCorr}, \Sec/ \ref{sec:DDEFacetting}) will tend not to be
a convolution operator, (ii) it decomposes the
signal in terms of Taylor basis functions, and the signal needs to be gridded
$n_t$-times if $n_t$ is the number of Taylor terms, and (iii)
baseline-dependent averaging cannot be used with A-Projection (see
\Sec/ \ref{sec:BDA}).

Instead, we produce a $(n_{ch}\times n_{pix})$ spectral cube, the
dirty images of size $(n_{pix})$ being formed into the corresponding
$n_{ch}$ frequency chunks. The spectral cube then contains information
about the sky's spectral properties.  We present in this section two
wideband deconvolution algorithms that estimate flux densities as
well as the intrinsic spectral properties (taking into
account Jones matrices such as primary beam direction-time-frequency behavior). The
first uses a variation of the matching pursuit \CLEAN/ algorithm,
while the second uses a genetic algorithm.

\subsection{\PMP/ deconvolution}
\label{sec:PMPClean}

\def\L2Q{
\left\|
\DirtyVecNorm-\TransferFuncPhiPi\vec{\pi_i}
\right\|_{\bm{Q}}
}

\def\CostFunction{
\mathcal{C}_{0}\{\TransferFuncPhiPi\vec{\pi_i}|\DeltaDirtyVecNorm,\bm{Q}\}
}

\def\ResidVis{
\Vis-\bm{\mathcal{A}}\bm{\Pi}\widehat{\piGlobal}
}

\begin{algorithm}[t!]
\SetAlgoLined
 \KwData{$\DirtyVecNorm$, $t$, $\alpha$, $\bm{Q}$, $\Vis$}
 \KwResult{
The estimated model $\widehat{\vec{\pi}}$ of
the true sky $\vec{\pi}$ in the basis function $\mathcal{P}$.}
initialization: $\widehat{\vec{\pi}}=\vec{0}$\;
\tcc{Start $n_{Cycle}$ major cycles}
 \For{$i_{Cycle}$ in range($n_{Cycle}$)}{
\tcc{Start minor cycles}
 \While{$\max{\{\DeltaDirtyVecNorm\}}>t$}{
   Find location of brightest pixel: \hspace{3cm}$i=\argmax{}\left(\DeltaDirtyVecNorm\right)$\;
   Find locally best sky model centered on pixel $i$:
   $\widehat{\vec{\pi_{i}}}=\argmin{\vec{\pi_{i}}}\left(\CostFunction\right)\ \left[\textrm{s.t. }\mathcal{C}_{1}\{\vec{\pi_i}\}\right]$\;\label{step:argmin1} 

   Update sky model: \hspace{3cm} $\widehat{\vec{\pi}}\leftarrow
   \widehat{\vec{\pi}}+\alpha\ \widehat{\vec{\pi_i}}$\;
   Update residual image: \hspace{3cm} $\DeltaDirtyVecNorm\leftarrow\DeltaDirtyVecNorm-\alpha\TransferFuncPhiPi\widehat{\vec{\pi_i}}$\;
   
 }
 Update residual image: \hspace{3cm} $\DeltaDirtyVecNorm =\widetilde{\MuellerImage}^{-H}\bm{\mathcal{A}}^H
 \bm{\mathcal{W}} \left(\ResidVis\right)$\;
}
Compute model image
 $\widehat{\vec{x_\nu}}=\bm{\Pi_\nu}\widehat{\vec{\pi}}$\;
Compute restored image
 $\widehat{\vec{x}}\leftarrow\DeltaDirtyVecNorm+\bm{\mathcal{C}}\widehat{\vec{x_\nu}}$\;

 \caption{\label{algo:HMP}\PMP/ deconvolution algorithm. Here $t$ is a
 user defined flux density threshold, $\alpha$ is a minor cycle loop
 gain. Other symbols are defined in Table~1 and/or in the main text.}
\end{algorithm}

In this section we present the \PMP/ deconvolution algorithm
(Hybrid Matching Pursuit). The idea is quite simple and
consists of decomposing the signal around the brightest pixel $i$ in
the spectral cube $\DirtyVecNorm$
into a sum of components with different spatial and spectral
properties. The basis function is similar to 
\MTMSCLEAN/ \citep{Rau11}, but the idea differs in that (i) we grid the data only once (we do not create dirty images at different
resolutions and for different Taylor terms), (ii)
the optimisation step is done on a set of pixels (and not only on the
brightest pixel), and
(iii) at each iteration all coefficients are estimated in the chosen
basis function (as opposed to the maximum coefficient only). This
last point is illustrated by the example of a faint extended signal
containing a brighter point source. While \citet{Cornwell08_MS}
have to introduce an {\it ad hoc} ``small-scale bias'' to reconstruct the compact emission, we aim at finding nonzero coefficients for the point source
and the extended emission, at each iteration (the same applies to the
spectral axis). The following algorithm
is implemented in \DDFacet/, natively taking direction-dependent
residual images and associated PSFs into account (see
\Sec/ \ref{sec:DDEFacetting_IMCORR} for a discussion of the
normalization).

We first choose a set 
$\mathcal{P}$ of functions into which we want to decompose the spectral
cube. For example, it can be made of Gaussians with various sizes and spectral
indices. The sky image $\vec{x^i_\nu}$ of models centered on pixel $i$
at a frequency $\nu$ is
then written as

\begin{alignat}{2}
\vec{x_\nu}=& \displaystyle\sum_i \PiLocalChan \piAtPixel
\eqdefT{\bm{\Pi_\nu}}\bm{\Pi_\nu}\piGlobal\\
\textrm{and }\vec{x}=&
\begin{bmatrix}         
\vdots\\
\vec{x_\nu}\\
\vdots\\
\end{bmatrix}
\eqdefT{\bm{\Pi}}\bm{\Pi}\piGlobal
\end{alignat}

\noindent where $\PiLocalChan$ is the $(\npix\times\npar)$ matrix
containing the spectro-spatial dictionary estimated at frequency $\nu$,
and $\vec{\pi_{i}}$ is the spectro-spatial sky model 
of pixel $i$, containing the $\npar$ parameter values of the
spectro-spatial dictionary. We can then write the contribution $\DirtyVecNormPixel$ of
pixel $i$ to the spectral cube as

\begin{alignat}{2}
\label{eq:ParmToDataSSD}
\DirtyVecNormPixel =& 
\begin{bmatrix}         
\vdots\\
\DirtyVecNormChanPixel\\
\vdots\\
\end{bmatrix}
=
\begin{bmatrix}         
\vdots\\
\TransferFuncDirNorm\PiLocalChan\\
\vdots\\
\end{bmatrix}
\piAtPixel\\
&\eqdefT{\TransferFuncPhiPi}\TransferFuncPhiPi\piAtPixel\\
\textrm{and }\DirtyVecNorm =
& \displaystyle\sum_{i}
\TransferFuncPhiPi\piAtPixel\\
&\eqdefT{\TransferFuncPhi}\TransferFuncPhi\piGlobal
\label{eq:GlobalTFunc}
\end{alignat}

\noindent where $\nu$ is the frequency chunk and
$\TransferFuncDirNorm$ is the normalized spectral PSF. In short, $\TransferFuncPhi$ maps the vector of
spatio-spectral coefiscients $\piGlobal$ for all pixels to the spectral cube
$\DirtyVecNorm$, taking into account the local spectral PSFs.

The algorithm is described in detail in
Alg. \ref{algo:HMP}. Particular attention needs to be given to Step
\ref{step:argmin1} where we estimate the best local model by
minimizing a cost function. Different cost functions give different
variations of \PMP/. Relaxing the constraint $\mathcal{C}_{1}$, we can for instance set $\CostFunction$ as

\begin{alignat}{2}
\label{eq:SSDL2}
\CostFunction&\eqdef\L2Q
\end{alignat}

\noindent where the $\mathcal{L}_2$ norm
$\left\|\vec{x}\right\|_{\bm{Q}}=\vec{x}^T\bm{Q}^{-1}\vec{x}$ of
$\vec{x}$ is computed for the metric $\bm{Q}$, with $\bm{Q}$ being in practice a
tapering function. The least-squares solution is then given by the
pseudo-inverse

\begin{alignat}{2}
\widehat{\piAtPixel}&=\left[\TransferFuncPhiPi^T\bm{Q}\ \TransferFuncPhiPi\right]^{-1}\TransferFuncPhiPi^T\bm{Q}\ \DirtyVecNorm
\end{alignat}

Alternatively, we can use a Non-Negative Least Squares (NNLS) optimisation in
Step \ref{step:argmin1} (Alg. \ref{algo:SSD}) by setting
$\CostFunction$ as in Eq.  \ref{eq:SSDL2} while constraining the
solution using
$\mathcal{C}_{1}\{\vec{x}\}\eqdef\left(\vec{x}>0\right)$. In our
experience the
\PMP/-{\sc nnls} gives the best results in reconstructing extended emission.

\subsection{Wide-band joint subspace deconvolution}
\label{sec:SecSSD}

In this section, we describe the \SSD/ (SubSpace Deconvolution
algorithm). It is a generic hybrid joint deconvolution algorithm that
uses subspace optimisation. We present in \Sec/ \ref{sec:SSD} the
generic scheme for subspace optimisation in the framework of
deconvolution, and in \Sec/ \ref{sec:SSD_GA} we present one such algorithm
that uses a genetic algorithm in the
optimisation step (\SSDGA/).

\subsubsection{Subspace optimisation for deconvolution}
\label{sec:SSD}

It is well known that deconvolution algorithms based on Matching-Pursuit solvers
(specifically \CLEAN/) are not robust in the deconvolution of
extended emission. Joint deconvolution algorithms are more robust, as
demonstrated by \citet[][]{Garsden15}, but are not useful with large
images since their sizes can exceed $\left(10^{4-4.5}\times10^{4-4.5}\right)$ pixels. 
Indeed, Eq. \ref{eq:GlobalTFunc} is costly to invert because
$\TransferFuncPhi$ is expensive to
apply\footnote{$\TransferFuncPhi$ is done computationally by {\it
    degridding} the data}. Therefore in order to make joint
deconvolution practical with real life data-sets, we aim at
incorporating it in a matching pursuit-type scheme.
As for \PMP/ (\Sec/ \ref{sec:PMPClean}), the idea is to
decompose the signal into a basis function but here the parameter space
at each iteration is not a set of coefficients for one pixel only,
but for a subset $\mathcal{I}$ of pixels in the spectral cube (an {\it
  island}). 


To illustrate the idea of \SSD/, consider the global transfer function
in Eq. \ref{eq:GlobalTFunc}. Since the convolution matrix is
diagonally dominant (the PSF goes to zero far from the center), the
main idea is that distant regions can be deconvolved separately. This
amounts to building an operator $\widetilde{\TransferFuncPhi}$ with zeros
where $\TransferFuncPhi$ is considered to be negligible such that
$\DirtyVecNorm\approx\widetilde{\TransferFuncPhi}\vec{\pi}$, and the
deconvolution is done jointly within each subspace $\{\piAtIsland\}$
of the global $\{\piGlobal\}$ parameter space.  This approximation
will however lead to biases in the estimate $\widehat{\vec{\pi}}$ of
$\vec{\pi}$, because the contribution of the sky in island
$\mathcal{I}$ to the observed flux in island $\mathcal{I}'$ has been
neglected. 

This will happen for example when a bright (a) source in an
$\mathcal{I}_a$ island has a faint (b) source ($\mathcal{I}_b$ island)
in its side-lobe, and when the two islands are deconvolved
independently. The faint source flux can be over- (or under-)
estimated in the first iteration since the cross-contamination term
is ignored. However if one computes the global residual map in a second
iteration, most of the side-lobe of source (a) has been properly
removed at $\mathcal{I}_b$ . If the islands are jointly deconvolved
again, the sky model estimate will be better than in the previous
iteration. In our experience, this algorithm has remarkable
convergence properties.


The \SSD/ algorithm is described in detail in
Alg. \ref{algo:SSD}. Given a residual image, in a first step the
brightest regions are isolated and joint deconvolution is performed
{\it independently} on groups of pixels (here called {\it islands})
using the local convolution operator
$\widetilde{\TransferFuncPhiPiIsland}$ with 
$\widetilde{\TransferFuncPhiPiIsland}=\widetilde{\TransferFunc^{\phi}}\bm{\mathcal{S}_{\scriptscriptstyle\mathcal{I}}}\bm{\Pi_{\scriptscriptstyle\mathcal{I}}}$,
where $\bm{\mathcal{S}_{\scriptscriptstyle\mathcal{I}}}$ is an
$(n_{pix}\times n^{\scriptscriptstyle\mathcal{I}}_{pix})$ matrix that maps
the $n^{\scriptscriptstyle\mathcal{I}}_{pix}$ pixels of island
$\mathcal{I}$ onto the full set of $n_{pix}$ pixels. For example we can minimize the cost
function by setting 

\begin{alignat}{2}
\mathcal{C}_{0}\eqdef
\left\|\DeltaDirtyVecNorm-
\widetilde{\TransferFuncPhiPiIsland}
\vec{d\pi_{\scriptscriptstyle\mathcal{I}}}
\right\|_{\bm{Q}}
\end{alignat}

\noindent where $\vec{d\pi_{\scriptscriptstyle\mathcal{I}}}$ are the
differential values of the spatio-spectral coefficients in a given basic function
(see \Sec/ \ref{sec:PMPClean}).

In a second step, the union of
the sky models are subtracted from the visibilities, and the visibilities are
re-imaged (corresponding to the step
$\DeltaDirtyVecNorm=\DirtyVecNorm-\TransferFuncPhi\widehat{\vec{\pi}}$).

\def\ErrorResid{\DeltaDirtyVec_{\overline{\mathcal{I}},k}}
\def\AbsErrorResid{\Abs{\ErrorResid}}

The conditions for the convergence of \SSD/ are hard to find, but depend on the
structure of $\widetilde{\TransferFuncPhi}$ compared to
$\TransferFuncPhi$. We can estimate at step $k$ the contribution to the
observed flux $\ErrorResid$ in $\mathcal{I}$ of all islands
$\mathcal{I'}\neq\mathcal{I}$. If
$\delta\widehat{\vec{x}}_{k}=\vec{x}-\widehat{\vec{x}}_{k}$ is the
error in the estimate $\widehat{\vec{x}}$ of $\vec{x}$, we can write

\def\DeltaTransFunc{\bm{\Delta_{\scriptscriptstyle\mathcal{I}}}}

\begin{alignat}{2}
\ErrorResid=& 
\bm{\mathcal{S}_{\scriptscriptstyle\mathcal{I}}}^T
\left(
\displaystyle\sum_{\mathcal{I'}\neq\mathcal{I}}
\widetilde{\TransferFunc}
\bm{\mathcal{S}_{\scriptscriptstyle\mathcal{I'}}}\bm{\mathcal{S}_{\scriptscriptstyle\mathcal{I'}}}^T
\right)
\delta\widehat{\vec{x}}_{k}
\end{alignat}

\noindent Since each island $\mathcal{I}$ is deconvolved in its own
subspace (independently of other islands), the level of the flux
density bias at iteration $k+1$ is

\begin{alignat}{2}
\Abs{\delta\widehat{\vec{x}}_{\scriptscriptstyle\mathcal{I},k+1}}\sim&\AbsErrorResid\\
&=\sqrt{\ErrorResid^T\ErrorResid}
\end{alignat}

\noindent Assuming the structures of the side-lobes of the different 
$\mathcal{I'}$ in $\mathcal{I}$ are uncorrelated, the power in the cross-island terms averages out in the
quadratic sum, and we get

\def\GainPIsland{\bm{\mathcal{G}_{\scriptscriptstyle\mathcal{I}}}}

\begin{alignat}{2}
\Abs{\delta\widehat{\vec{x}}_{\scriptscriptstyle\mathcal{I},k+1}}&\sim
\sqrt{
\delta\widehat{\vec{x}}_{k}^T
\displaystyle\sum_{\mathcal{I'}\neq\mathcal{I}}
\left(
\bm{\mathcal{S}_{\scriptscriptstyle\mathcal{I'}}}^T
\widetilde{\TransferFunc}^T
\bm{\mathcal{S}_{\scriptscriptstyle\mathcal{I}}}
\bm{\mathcal{S}_{\scriptscriptstyle\mathcal{I}}}^T
\widetilde{\TransferFunc}
\bm{\mathcal{S}_{\scriptscriptstyle\mathcal{I'}}}
\right)
\delta\widehat{\vec{x}}_{k}
}\\
&\eqdefT{\GainPIsland}
\sqrt{
\delta\widehat{\vec{x}}_{k}^T
\GainPIsland
\delta\widehat{\vec{x}}_{k}
}
\end{alignat}

\noindent Here $\GainPIsland$ is the power in the side-lobes
of all islands $\mathcal{I'}$ to islands $\mathcal{I}$. If the
cross-contamination power is small enough \SSD/ converges. For example, in
the trivial case of two single pixel islands with equal flux $s$, and
cross-contamination term $p$ (the PSF of $\mathcal{I}$ onto
$\mathcal{I'}$ and conversely), at iteration $k$ we have

\begin{alignat}{2}
\Abs{\delta\widehat{\vec{x}}_{\scriptscriptstyle\mathcal{I},k+1}}&\sim
\Abs{\delta\widehat{\vec{x}}_{\scriptscriptstyle\mathcal{I'},k}}
\sqrt{p^2}\\
&\sim s \sqrt{g^2}^{k+1}
\end{alignat}

\noindent and SSD always converges.

\def\CostFunction{
\mathcal{C}_{0}\{
\widetilde{\TransferFuncPhiPiIsland}
\vec{d\pi_{\scriptscriptstyle\mathcal{I}}}
|\DeltaDirtyVecNorm,\bm{Q}\}
}

\def\ComplicatedLine{
$\widehat{\vec{d\pi}_{\scriptscriptstyle\mathcal{I}}}=\argmin{\vec{d\pi_{\scriptscriptstyle\mathcal{I}}}}
\left(
\CostFunction
\right)
\left[\textrm{s.t. }\mathcal{C}_{1}\{\vec{d\pi_{\scriptscriptstyle\mathcal{I}}}\}\right]$
}

\begin{algorithm}[t!]
\SetAlgoLined
 \KwData{$\DirtyVecNorm$, $t$, $\alpha$, $\bm{Q}$}
 \KwResult{
The estimate $\widehat{\vec{\pi}}$ of
$\vec{\pi}$ in the basis function $\mathcal{P}$.}
initialization: $\widehat{\vec{\pi}}\leftarrow\vec{0}$,$\DeltaDirtyVecNorm\leftarrow\DirtyVecNorm$\;
\tcc{Start $n_{Cycle}$ major cycles}
 \While{$\max{\{\DirtyVecNorm\}}>t$}{
\tcc{Deconvolve each island $\mathcal{I}$ independently}
 \For{$\mathcal{I}$ in $\{\mathcal{I}\}$ }{
    \ComplicatedLine\;
 \label{step:argminSSD} 
}
   
   Update sky model: \hspace{3cm} $\widehat{\vec{\pi}}\leftarrow
   \widehat{\vec{\pi}}+\displaystyle\sum_{\mathcal{I}}\bm{\mathcal{S}_{\scriptscriptstyle\mathcal{I}}}\widehat{\vec{d\pi}_{\scriptscriptstyle\mathcal{I}}}$\;
Compute the residual data:
$\DeltaDirtyVecNorm =\widetilde{\MuellerImage}^{-H}\bm{\mathcal{A}}^H
 \bm{\mathcal{W}} \left(\ResidVis\right)\left(=\DirtyVecNorm-\TransferFuncPhi\vec{\widehat{\pi}}\right)$\;
   
}

 \caption{\label{algo:SSD}\SSD/ deconvolution algorithm. Here $t$ is a
 user defined flux density threshold.}
\end{algorithm}

\subsubsection{An example of \SSD/ using genetic algorithm}
\label{sec:SSD_GA}

We have presented in \Sec/ \ref{sec:SSD} the \SSD/ algorithm, which
carries out
joint deconvolution over a set of sub-spaces in an independent
manner. In this section we detail how the genetic algorithm in \SSDGA/
implements step \ref{step:argminSSD} in
Algo. \ref{algo:SSD}. Specifically, we discuss an example of an \SSD/
algorithm, where we perform step \ref{step:argminSSD}
(Algo. \ref{algo:SSD}) using a genetic algorithm (\SSDGA/). Genetic
algorithms are very different from convex solvers in the sense that
they are (i) combinatorial and (ii) nondeterministic. While genetic
algorithms are rather simple to use and very flexible, \SSDGA/ is in
principle good for the deconvolution of extended signal. We can for
instance optimize the $\mathcal{L}_0$ norm which is a nonconvex
problem.

This step corresponds to fitting the residual dirty image by a
spectral sky-model for each island $\mathcal{I}$, convolved by the
local spectral PSF.

Our current implementation is based on the {\sc deap} package
\citep{DEAP}. Each individual `sourcekin' consists of a set of fluxes
together with a spectral index. Each sourcekin is a spectro-spatial
model of the sky in $\mathcal{I}$. It could also include minor axis,
major axis, and position angle of a Gaussian for example. The idea
consists of building and evolving the population of sourcekin, and the
fitness function is set to be $\mathcal{L}_2$ in our case. An example
of spectral deconvolution using \SSDGA/ is presented in
\Sec/ \ref{sec:Simulation}.

\def\FigWidth{4}
\begin{figure}[]
\begin{center}
\includegraphics[width=\FigWidth cm]{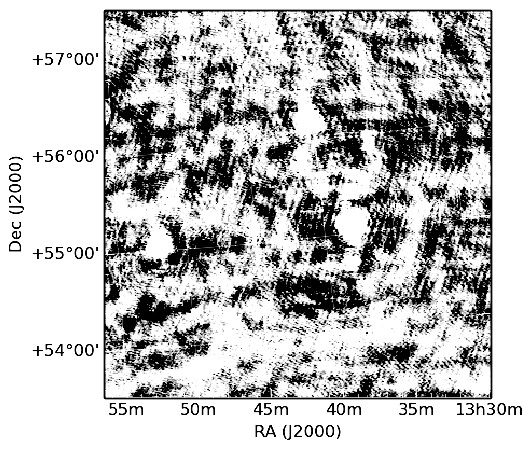}
\includegraphics[width=\FigWidth cm]{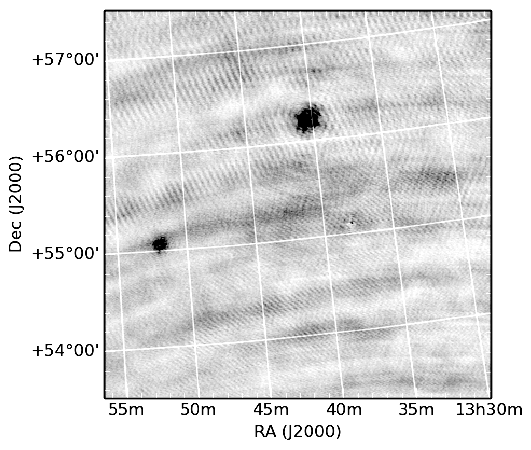}
\includegraphics[width=\FigWidth cm]{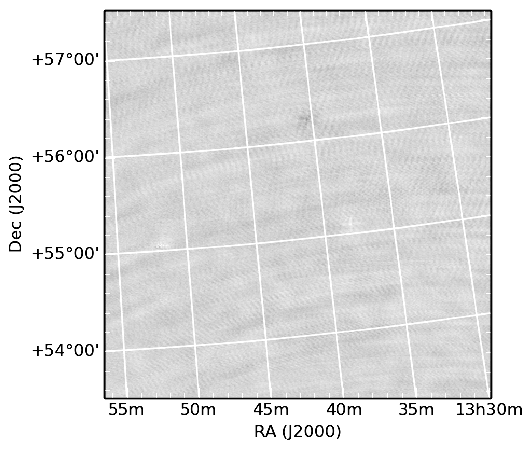}
\includegraphics[width=\FigWidth cm]{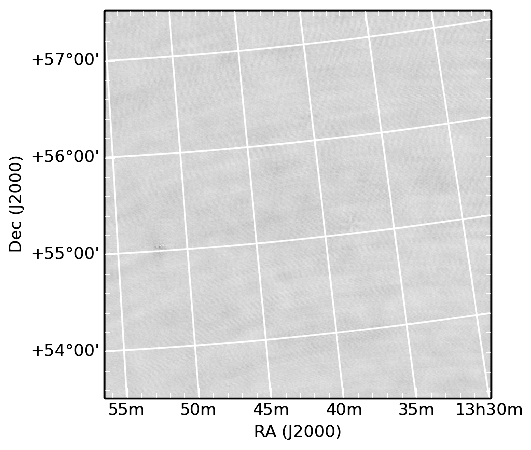}
\caption{\label{fig:SSD_CONV} This figure shows the residual data for
  a fraction of the field of view of the simulation described in
  \Sec/ \ref{sec:Simulation}. The left to right and top to bottom
  panels show the iterations $\{0,1,2,3\}$. As explained in \Sec/ \ref{sec:SSD}, the \SSD/ algorithm works differently
  from a matching pursuit in that it does joint deconvolution on
  subsets of pixels, and the estimated flux is fully removed at
  each iteration. The \SSD/ has remarkable convergence properties.}
\end{center}
\end{figure}

\section{Implementation, performance and features}

\label{sec:Implementation}

The bulk of \DDFacet/ is implemented in Python 2.7, with a small performance-critical core 
module (gridding and degridding) written in C. In this section we discuss some important aspects of the
implementation. In \Sec/ \ref{sec:parallel}, we describe aspects of parallelization. In \Sec/ \ref{sec:BDA}, 
we describe how we use a baseline-dependent averaging scheme in the context of wide-field wideband
spectral deconvolution, and we explain how we handle the nonregular
spatial domains of Jones matrices in \Sec/ \ref{sec:tessel}. In
\Sec/ \ref{sec:Simulation} we demonstrate our imaging and deconvolution
framework on a single simulation.

\subsection{Parallelization}
\label{sec:parallel}

The gridding, degridding and FFT operations of faceted imaging are embarrassingly parallel, as every facet 
can be processed completely independently. The \DDFacet/ implementation is parallelized at the single-node level, using the
Python {\tt multiprocessing} package for process-level parallelism, and a custom-developed process manager called
{\tt AsyncProcessPool} that implements asynchronous, on-demand, multiprocessing akin to the concurrent 
futures\footnote{\url{https://docs.python.org/3.4/library/ concurrent.futures.html}} module found in Python 3. 
The bulk of the data (visibilities, $uv$-grids and and images) is stored in shared memory using the 
SharedArray\footnote{\url{https://pypi.python.org/pypi/SharedArray}} module, and a custom extension 
called {\tt shared\_dict}. This significantly reduces the overhead of inter-process communication. This also allows 
us to perform I/O and computation concurrently: a successive data chunk is read in while gridding of the
previous chunk proceeds. In the minor cycle, we employ the same technique to parallelize the \SSD/ algorithm. For \PMP/ deconvolution
(and other CLEAN-style minor loops), the minor loop is inherently serial, but a reasonable speedup is achieved 
with minimum effort by employing the {\tt numexpr} package\footnote{\url{https://pypi.python.org/pypi/numexpr}} to vectorize
large array operations.

All this allows \DDFacet/ to make very good use of multiple cores in a NUMA architecture, maintaining 
 high core occupancy throughout any given imaging run. We are conducting detailed performance studies and these
will be the subject of a separate paper. Here we present the summary results of a simple parallelisation scaling experiment.

We perform an imaging run using 14 hours of VLA (C+D configuration) data for the field around the source 
3C147, in L-band. This totals 2350127 time-baseline samples, with 64 channels each, for a total bandwidth 
of 256 MHz. We make $5100\times5100$ pixel images of a $2.8^\circ\times2.8^\circ$ field tiled by $23\times23=529$ square 
facets, in two frequency bands of 128 MHz each. A (rotating) primary beam model is applied on a per-facet basis. 
We run 5 major cycles of \PMP/ CLEAN, down to an absolute flux threshold of 0.4 mJy. 

\begin{figure}[]
\begin{center}
\includegraphics[width=\columnwidth]{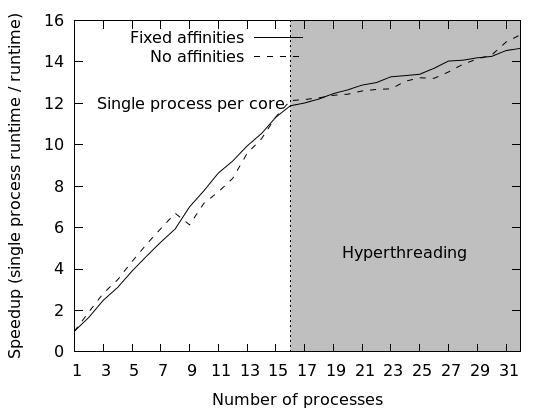}
\caption{\label{fig:speedups}The speedup factor (in terms of overall wall time) obtained by running \DDFacet/ on 
multiple cores. The solid line corresponds to fixed affinities (each worker process was pinned to a 
single CPU core), while the dashed line to no affinities (the OS scheduler was allowed to migrate processes 
across cores). In the former case, processes 0--7 were pinned to the first physical CPU, and 8--15 to the 
second CPU. This explains the slightly better performance in the 
no-affinities regime with $\leq 8$ processes, as the OS scheduler was allowed to make use of the second CPU.
The graph also shows significanly worse scaling in the hyperthreaded regime.}
\end{center}
\end{figure}

Our test machine has two Intel Xeon E5, Sandy Bridge class CPUs, each with 8 physical cores and 16 virtual cores 
(hyperthreading enabled). In serial mode, i.e. with all operations running on a single core, we measure a 
total ``wall time'' for this imaging run of about 12 hours. 94\% of this time is spent in the gridding. 
We then increase the
number of parallel processes, and plot the resulting speedup factor (in terms of wall time, thus including all
overheads) in Fig.~\ref{fig:speedups}. We see exemplary linear scaling of performance up to 16 processes (i.e. to the 
point where each physical core is occupied by a single process). Beyond this point, the scaling relation declines, 
as processes running on virtual cores start competing for resources of a single physical core. 
Note that a speedup factor of 12 from 16 cores is excellent efficiency: a quick calculation shows that 
this corresponds to 98\% of the computation being parallelized.

From this we can conclude that our parallel implementation scales linearly with available physical cores, while 
the benefits of hyperthreading are marginal in comparison. We also find that the computational cost of the gridding step dominates overall processing. 
\DDFacet/ therefore implements two strategies for reducing the overall cost of gridding: \emph{baseline-dependent averaging} (BDA) and
\emph{sparsification}. 

\subsection{Baseline-dependent averaging}
\label{sec:BDA}

Averaging visibilities has the effect of reducing data volumes and
increasing computing efficiency. However, information is unavoidably lost in the process, and
therefore inverting the Measurement Equation from the averaged (and
therefore {\it smaller}) set of visibility measurements is, numerically, subject 
to poorer conditioning. 

The metric we use to limit the loss of information is based on
decorrelation effects\footnote{Decorrelation is produced by averaging
of complex-valued Jones matrices. It is easy to see, for instance,
that averaging a rotating complex scalar will result in a loss of
amplitude.}, and those will indeed constrain the maximum time
and frequency domain over which visibilities can be averaged. It can
be seen from the \RIME/ (Eq. \ref{eq:ME}) that decorrelation can be caused by the
variation over time, frequency and direction of (i) the Jones matrices
or (ii) the sky, and most importantly (iii) the geometric phase term
(the {\it kernel} $k$ term in Eq. \ref{eq:ME}). Decorrelation due (i)
and (ii) largely depends on the target, the instrument and the
observing frequency. For example, low-frequency $\nu\lesssim300$
MHz data (such as that taken by the LOFAR telescope) 
is affected by ionospheric phase, which varies
on the timescale of $10\sim30$ seconds (and is also direction-dependent
due to the large FoVs). At higher frequencies, tropospheric phase begins 
to have a similar (although effectively direction-independent) effect.
The decorrelation due to (iii) is well understood and predictable.
For a given direction $\vec{s}$, if the phase
varies linearly across the time or frequency domain $\mathcal{D}$, and
one can write

\def\DomainPQ{\scriptscriptstyle\mathcal{D}}

\begin{alignat}{2}
\chi_{\DomainPQ}=&\sinc{\Delta\psi_{\DomainPQ}}\\
\textrm{with }\Delta\psi_{\DomainPQ}=&\pi\delta \vec{b}^T_{\DomainPQ}\vec{s}\\
\textrm{and }\delta \vec{b}_{\DomainPQ}=&\vec{b}_{\DomainPQ_1}-\vec{b}_{\DomainPQ_0},
\end{alignat}

\noindent where $\vec{b}_{\DomainPQ_0}$ and $\vec{b}_{\DomainPQ_1}$ are the
baseline vectors (in units of wavelength) at the edges of the domain $\mathcal{D}$.

Interferometric data is typically conservatively averaged at best, using a common time-frequency bin
across {\it all} baselines that corresponds to no more than a few percent amplitude loss on the 
{\it longest} baseline for a source on the edge of the field of view.
Several authors have come to the conclusion 
that this is sub-optimal, and that one could use baseline-dependent averaging (BDA) instead 
\citep[see][]{cotton1989special,cotton1999special,atemkeng2016}, with more agressive averaging on the 
shorter baselines, since for a given direction, time and frequency domain, they decorrelate less than 
the longer baselines. With core-heavy arrays such as MeerKAT and SKA1, the potential storage 
savings of BDA can be substantial, since the data sets are dominated by short spacings. 

It is important to keep in mind that, for purposes of data \emph{storage}, the largest time/frequency 
domain to which any given baseline may be averaged is given by

\def\DomainK{\mathcal{D}_{k}}
\def\DomainSky{\mathcal{D}_{\mathrm{s}}}
\def\DomainJ{\mathcal{D}_{\mathrm{J}}}

\begin{alignat}{2}
\mathcal{D}=&\textrm{min}\left\{\DomainJ,\DomainSky,\DomainK\right\},
\end{alignat}

\noindent where $\DomainJ$, $\DomainSky$, and $\DomainK$ are the
domains corresponding to an acceptable decorrelation for (i) the Jones
matrices, (ii) the sky, and (iii) the geometric phase term
respectively. In the presence of DDEs such as the ionosphere,
$\DomainJ$ is the term that typically constrains $\mathcal{D}$, because
there is no way to correct the stored visibilities for the DDEs; rather, 
we need to {\it apply} Jones matrices to the data during imaging, as
described in \Sec/ \ref{sec:DDEFacetting} (see Eq. \ref{eq:ME_LinAlg}
and \ref{eq:TransferFunc}). For example, consider a single
baseline at low frequency, having a decorrelation time scale
of the geometric phase term of the order of a few minutes. As the
Jones matrices corresponding to the ionosphere are direction
dependent, and vary on an approximately ten second time scale, one cannot average
the stored data on time scales larger than seconds without substantially 
degrading the imaging response.

Even if the storage economies of BDA are not realized (and are in any case limited
by $\DomainJ$), \DDFacet/ derives computational economies from this technique. If one assumes that the sky term is constant across $\DomainK$, one can average the $2\times2$
visibilities over $\DomainK$, while applying the per-facet phase rotations, and the (direction-dependent) 
Jones matrices, to each individual visibility. This is done on-the-fly in the gridder and degridder code. 
The actual gridding (or degridding) is then done once per averaging domain, rather than once 
per visibility. The resulting savings can be substantial, since
averaging visibilities involves fewer FLOPS than applying the convolution kernel inherent to gridding or degridding.
The BDA gridding algorithm implemented in \DDFacet/ is presented in detail in Algo. \ref{algo:Gridding}. We note
that a similar approach has been implemented in recent versions of WSCLEAN \citep{wsclean}, without the 
direction-dependent Jones correction.

\begin{figure}[]
\begin{center}
\includegraphics[width=\columnwidth]{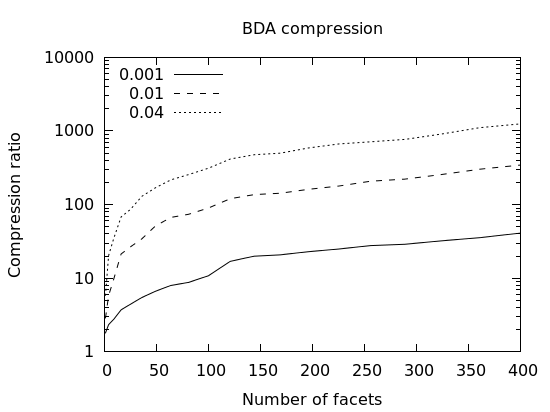}
\caption{\label{fig:bda-comp}Compression ratios achieved with
  baseline-dependent averaging, as a function of facet number (thus
  facet size), for several different decorrelation levels, for a VLA B-configuration
  observation.}
\end{center}
\end{figure}

\begin{figure}[]
\begin{center}
\includegraphics[width=\columnwidth]{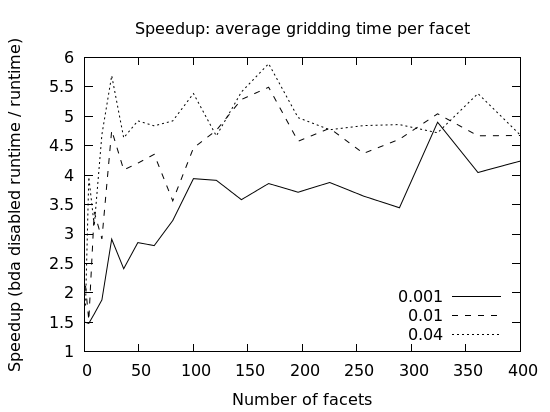}
\caption{\label{fig:bda-speedup}Gridding speedup factors achieved with baseline-dependent averaging, as a function of facet number (thus facet size), for several different decorrelation levels. Observation configuration as per previous figure.}
\end{center}
\end{figure}

On-the-fly BDA in the context of faceting offers an interesting performance trade-off. Note that 
$\DomainK$ is determined by facet size, rather than the full FoV size. While imaging smaller facets, more agressive 
BDA may be applied, since more visibilities can be averaged before a given decorrelation level is reached.
Note that at the limit of single-pixel facets, per-facet BDA reduces to averaging 
the entire (phase-shifted) dataset, which is effectively the same as doing a DFT. Figure~\ref{fig:bda-comp} shows 
the compression ratio achieved for a few fixed decorrelation levels, as a function of number of facets 
(across the same FoV), for a VLA B-configuration obervation. Note that more core-heavy
configurations such as MeerKAT and SKA1-MID should be able to achieve even higher compression ratios.
Figure~\ref{fig:bda-speedup} shows the resulting speedup in gridding time per facet. Note that the speedup flattens 
out at around a factor of 4. Presumably, at this point the gridder performance becomes dominated by memory access.
Thus, the computational cost of using numerous smaller facets (resulting in more gridding/FFT operations) is 
partially offset by the computational savings of increased BDA within each facet.

\def\VDomain{\bm{\mathrm{V}}_{\mathcal{D}}}
\def\VDomainVec{\bm{\mathrm{v}}_{\scriptscriptstyle\mathcal{D}}}
\def\Grid{\bm{g}_{\vec{s}_{\varphi}}}

\begin{algorithm}[t!]
\SetAlgoLined
 \KwData{$\Vis$, $\{\mathcal{D}\}$}
 \KwResult{
   The dirty cube $\DirtyVec,\DirtyVecNorm$\;}
$\vec{y}=\vec{0}$\;
 \For{$\varphi$ in $\{\varphi\}$ }{
 $\Grid=\vec{0} ; w_t=0$\;
 $\widetilde{\MuellerImageDir}=\vec{0}$\;
   \For{$\mathcal{D}$ in $\{\mathcal{D}\}$ }{
     $\VDomain=\bm{0} ; \vec{b}_m=\left(\u_{m},\v_{m},\w_{m}\right)=\vec{0}$\;
     \For{$\left(\vec{b},w,\JonesMat_{p},\JonesMat_{q},\Vis_{pq}\right)$ in $\mathcal{D}$ }{
       $\vec{b}_{m}\leftarrow \vec{b}_{m}+\MatFacetCoord\vec{b}$ (see
       Eq. \ref{eq:CoordUVTranf})\;
       $w_{m}\leftarrow w_{m}+w$\;
          $\MuellerImageDir=\JonesMat_{q}^*\kron\JonesMat_{p}$\;
        $\widetilde{\MuellerImageDir}^2\leftarrow\widetilde{\MuellerImageDir}^2+w\MuellerImageDir^H\MuellerImageDir$\;
       $\VDomainVec\leftarrow\VDomainVec+\MuellerImageDir\Vis_{pq}\exp{\left(-2\pi i\vec{b}^T\vec{s_\varphi}\right)}$\;
     }
     $\vec{b}_{m}\leftarrow \vec{b}_{m}/w_{m}$\;
     $\Grid\leftarrow\Grid+\bm{\mathcal{C}}_{\vec{s_\varphi},\vec{b}_{m}}\textrm{Vec}\left\{\VDomain\right\}$\;
     $w_t\leftarrow w_t+w_m$\;
   }
     $\Grid\leftarrow\Grid/w_t$\;
     $\DirtyVecNorm\leftarrow \DirtyVecNorm + \bm{\mathcal{S}}_{\vec{s}_0}\widetilde{\MuellerImage_{\vec{s_\varphi}}}^{-H}\Fourier^H\Grid/w_t$\;
 }

 \caption{\label{algo:Gridding}\BDA/ gridder. Here $\Grid$ is the grid
 used for facet $\varphi$,
 $\bm{\mathcal{C}}_{\vec{s_\varphi},\vec{b}_{m}}$ is the convolution
 function for the w-coordinate of $\vec{b}_{m}$}
\end{algorithm}

\subsection{Sparsification}

Recent developments in compressive sensing (CS) theory \citep[see, e.g.,][and references therein]{SARA} have provided new 
mathematical insights into imaging and deconvolution. In particular, CS shows that \emph{sparse} signals, i.e. signals with 
limited support in some dictionary, can be successfully recovered from a much smaller number of measurements than that 
required by e.g. Nyquist sampling, provided a few mathematical criteria (in particular, the Restricted Isometry Property, or RIP) 
are met. Traditional CLEAN has been shown to be a variation of a CS algorithm.

When multiple major cycles are performed by a CLEAN-like algorithm (as is necessary for even modestly high dynamic range imaging), 
the models recovered during early major cycles tend to be extremely sparse. This is simply due to the inherent structure of 
the radio sky: bright sources are few and far between, so it is quite typical that early major cycles of CLEAN affect only 
a relatively small number of components. Under these conditions, it directly follows from CS theory that we can 
discard
a random subset of $uv$-samples in the early major cycles (the randomness of the selection ensures that the RIP is maintained), 
and still recover the same model components, as long as the image SNR remains sufficiently high. 

\DDFacet/ implements this idea as the \emph{sparsification} feature. The implementation naturally interfaces with the BDA gridder. 
At each major cycle, a user-specified percentage of visibility blocks (i.e. BDA blocks) is tagged as ``unlucky'' and omitted from
the gridding process entirely. Effectively, the $uv$-coverage is randomly sparsified by a given factor, without changing the
relative sampling density. The gridding loop is therefore accelerated by the same factor. The sparsification factor can be set to 
decrease in successive major cycles, since, at the very least, the final major cycle should be done on the full data.

What is a reasonable sparsification factor? Consider a typical VLA or MeerKAT L-band observation of several hours' duration. 
This will contain $10^6\sim10^7$ time-baseline samples (per each frequency channel), and offer an image dynamic range of, 
conservatively, $10^4$ (in the presence of very bright sources, this can 
go to $10^6$ and above). Assuming a major cycle gain of 0.1, the first major cycle will clean down to $1000\sigma$, and the 
second major cycle to $100\sigma$, where $\sigma$ is the noise rms. A sparsification factor of 100 will reduce SNR by a 
factor of 10, with a negligible effect on overall $uv$-coverage, and with the second major cycle remaining well above the noise 
floor. (In practice, we find that sparsifying above a factor of 100 is not really useful, since at that point the runtime becomes 
dominated by I/O rather than gridding.) For 5 major cycles, a typical list of sparsification factors would be 100, 100, 30, 10, 1. 
This results in very substantial acceleration of the first four major cycles.

\subsection{Nonregular tessellation}
\label{sec:tessel}

Since astrophysical flux density is not uniformly distributed over the sky, a reasonable DDE calibration strategy consists of 
clustering sets of sources together, and estimating time-frequency-antenna dependent Jones matrices in each of those 
directions. As shown by \cite{Williams16} and \cite{Weeren16}, it is necessary in that context to image
the residual data in the nonregular spatial domains within which the
Jones matrices have been estimated.

\DDFacet/ has the ability to image irregularly tessellated images while using 2-dimensional FFTs. This is done by
providing it with an arbitrary set of nodes (i.e. facet centers). A Voronoi tessellation is then computed internally to 
generate a polygon file, where each facet is associated with a unique polygon. An image mask is constructed for each 
polygon. In order to form the residual image, each facet image is multiplied by its corresponding mask, before being stacked 
onto the combined residual image. We do the converse operation for the degridding.

\subsection{Primary beam models}

The present implementation of \DDFacet/ can correct for two classes of DDEs: direction-dependent gain solutions derived from
a calibration procedure, and/or \emph{a priori} known DDEs imposed by the primary beam pattern.

The first class of DDEs is specified as an input list of directions and (frequency-dependent) per-antenna Jones terms associated 
with that direction. Facets are then determined by the nonregular tesselation procedure described above, and imaging proceeds
on a per-facet basis, by applying the correct DDE solution (and primary beam E-Jones, computed at facet center) per facet, as described in 
\Sec/ \ref{sec:DDEFacetting}. In the absence of DD gain solutions, imaging can be done with primary beam 
corrections alone. In this case the field of view is split into a user-supplied number of square facets, and the primary 
beam E-Jones is again computed at each facet's center.

Two primary beam modes are currently implemented. In {\tt FITS} mode, the user specifies the primary beam pattern as a set of 8
FITS cubes\footnote{In principle it should also be possible to specify a different pattern per antenna, although this
option is currently not exposed in the interface.} giving the real and imaginary components of the $2\times2$ E-Jones 
elements, as a function of frequency and direction in the antenna frame. \DDFacet/ then computes parallactic angle rotation, 
and interpolates the E-Jones value appropriate to each facet's center, as a function of time and frequency. FITS mode is suitable to 
describing dish arrays. In {\tt LOFAR} mode, the primary beam
corresponding to a LOFAR station is computed using the {\tt
  LofarStationResponse} class.

\section{Simulations}

\label{sec:Simulation}

\def\LBAINNER/{{\sc lba}\_{\sc inner}}

\def\FigWidth{5.5cm}
\begin{figure*}[]
\begin{center}
\includegraphics[width=\FigWidth]{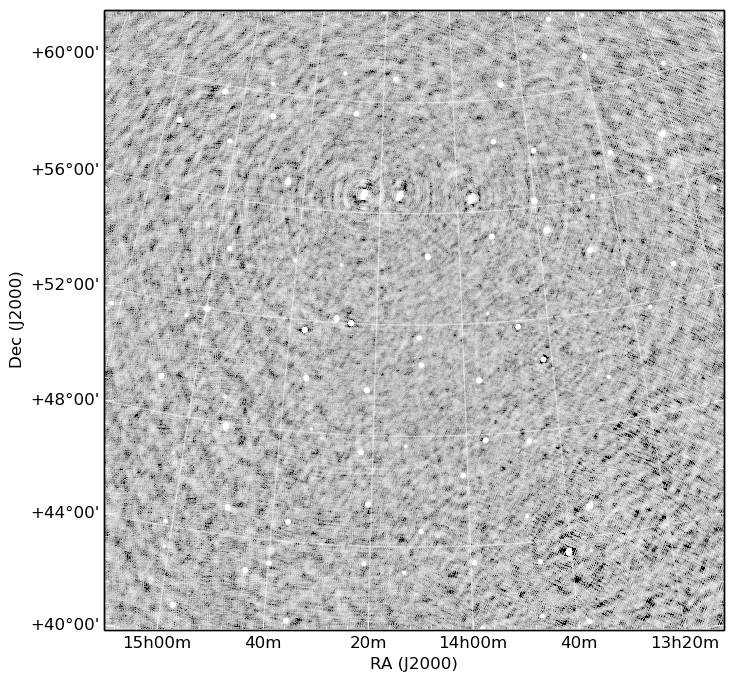}
\includegraphics[width=\FigWidth]{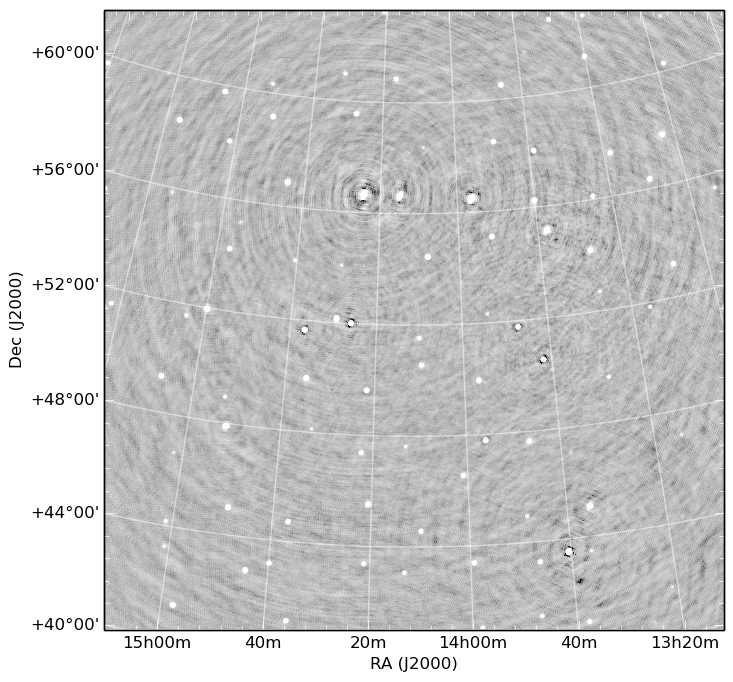}
\includegraphics[width=\FigWidth]{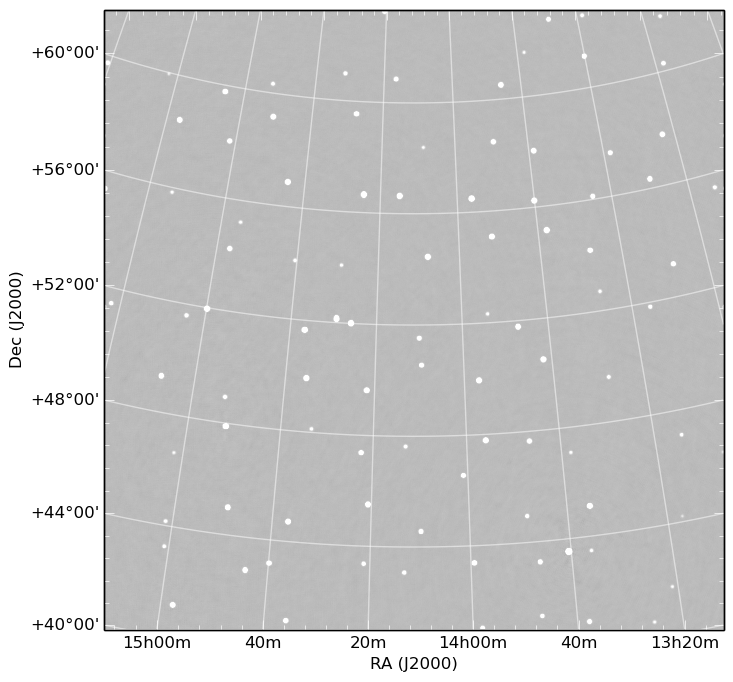}
\caption{\label{fig:Sim_Restored} This figure shows the deconvolved
  images for the simulated data-set described in
  \Sec/ \ref{sec:Simulation}. The left panel shows the restored
  image not taking the beam into account and not using wideband
  deconvolution. Central panel shows the result of the deconvolution
  taking DDE into account, but still not enabling wideband
  deconvolution. Right panel shows the deconvolution
  result using \SSDGA/.
}
\end{center}
\end{figure*}

\begin{figure*}[ht!]
\begin{center}
\includegraphics[width=\columnwidth]{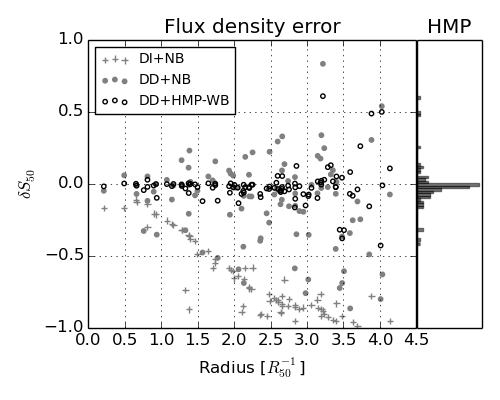}
\includegraphics[width=\columnwidth]{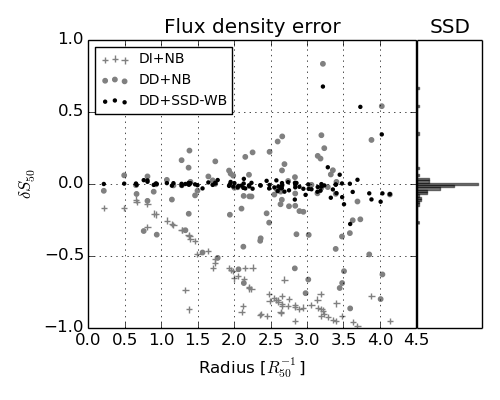}
\caption{\label{fig:Sim_Flux} This figure shows the relative flux
  density error 
$\delta S_{50}=\left(\widehat{S_{50}}-S_{50}\right)S^{-1}_{50}$ 
between the $50$ MHz flux density $S_{50}$ and its estimate
$\widehat{S_{50}}$ as a function of radius from the beam center
(normalized by the half power radius at $50$ MHz). The left and right panels show the results
for the \PMP/ and \SSD/ wide band
deconvolutions (empty and filled black circles respectively). The gray
cross corresponds to the flux densities errors taking neither DDE
correction nor wideband effects into account. The gray dots
show the results
when taking DDE effects, but not wideband effects into account.}
\end{center}
\end{figure*}

In this section we discuss in detail a test case for the framework developed
throughout this paper. We simulate a small, semi-realistic,
LOFAR LBA data-set. The data consists of an interferometric data-set
with $630$ baselines ($36$ LOFAR stations), $2.8$ hours long, with
$5$ frequency channels spanning
the range $30$ to $70$ MHz. The station are configured in \LBAINNER/ mode \citep{Haarlem13} in order to provide us with the widest field of
view. The HPBW is on the order $5.77$ degrees at $60$ MHz. 

The simulated sky consists of $\sim100$ point sources with $50$-MHz intrinsic
flux densities $S_{50}=s^{1.5}$ with $s\sim \mathcal{U}\left\{0,100\right\}$ Jy, and spectral indices
$\alpha_{50}\sim\mathcal{U}\left\{-1,1\right\}$, where $\mathcal{U}$ is the
uniform distribution. To make the case more difficult yet, we
add a bright $10^4$ Jy off-axis source.

Apart from the \LBAINNER/ station beam
being applied, scalar time-frequency-direction
dependent Jones matrices are applied to
the predicted visibilities. In particular, the Jones matrices have a random
phase and amplitude term that varies sinusoidally with time, and
this random realization is repeated for each antenna, frequency and direction. The typical amplitude variation is on the level of $0.1$,
while the phase variation is on the level of $0.1\pi$. This scheme guarantees (i) nonunitarity of the Jones matrices (and Mueller
matrices), and (ii) baseline-dependency of the resulting Mueller
matrices (see \Sec/ \ref{sec:DDEFacetting_IMCORR}).

The Jones matrices together with the LOFAR \LBAINNER/ beam are provided to \DDFacet/ and the sky is tessellated
as described in \Sec/ \ref{sec:tessel}. The image is $8019\times8019$
pixels with a pixel size of $10\times10$\arcsec , and when the wideband mode is enabled the residual image cube
has $3$ channels. 

To measure the estimated sky produced by the different deconvolution
algorithms, we run \DDFacet/ in various modes:
\begin{itemize}
\item[\textbullet]{{\sc di/dd}: Direction-independent faceting or
  direction-dependent faceting.}
\item[\textbullet]{{\sc nb/wb} Narrow-band and wideband observations. In the latter
  case the intrinsic spectral variability is taken into account
  either by the \PMP/ (\PMP/-{\sc wb}) or by the \SSDGA/ (\SSD/-{\sc wb}).}
\end{itemize}

\begin{figure}[h]
\begin{center}
\includegraphics[width=\columnwidth]{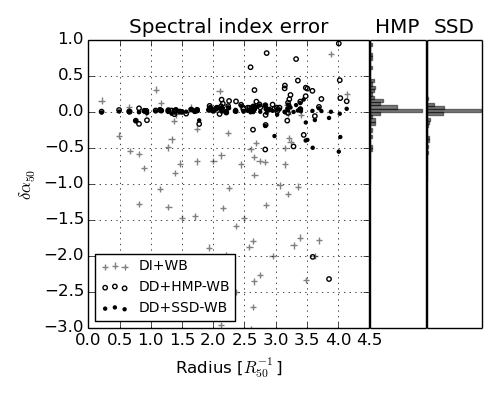}
\caption{\label{fig:Sim_Alpha} This figure shows the error 
$\delta\alpha_{50}=\widehat{\alpha_{50}}-\alpha_{50}$ between the
  spectral index $\alpha_{50}$ and its estimate
  $\widehat{\alpha_{50}}$, as a function of the distance from the
  beam center. The gray crosses show the spectral index estimates when
  the DDEs are not taken into account.}
\end{center}
\end{figure}

The results are presented in Figs. \ref{fig:Sim_Restored}, \ref{fig:Sim_Flux} and
\ref{fig:Sim_Alpha}. Fig.~\ref{fig:Sim_Restored} shows that, as expected, the residuals are much
lower when deconvolved with {\sc dd+nb} and {\sc di+wb}. In Fig. \ref{fig:Sim_Flux}, we plot the relative flux error 
$\delta S_{50}=\left(\widehat{S_{50}}-S_{50}\right)S^{-1}_{50}$ as a
function of the distance to the beam center. It is quite clear from
the plot that, as expected, the {\sc di+nb} gives biased flux density
estimates and higher residual values (see
Fig. \ref{fig:Sim_Restored}). The main component of the bias is
corrected using {\sc dd+nb}, but the flux density scatter is
high. With \PMP/ or \SSD/ deconvolution, the dynamic range increases, and the
error on both flux densities and spectral indices decreases. In
Fig. \ref{fig:Sim_Alpha}, we plot the error $\delta\alpha_{50}$ in
spectral index estimate as a function of the distance from the beam
center. While {\sc di+wb} produces biased spectral index estimates,
the \PMP/ or \SSD/ algorithms properly recover them (with \SSD/ doing slightly better).

\section{Overview and future work}
\label{sec:Conclusion}

In this paper we have presented a mathematical framework for wideband wide-field spectral
deconvolution that can take generic DDEs into account, as well as an implementation of this framework 
called the \DDFacet/ imager. This has a number of unique features not available in other imaging and deconvolution 
frameworks:

\begin{itemize}
\item[\textbullet]{A wide-field coplanar faceting scheme is
  implemented. This is a generalization of the \citet{Kogan09} scheme.
Nontrivial facet-dependent $w$-kernels are used to correct for noncoplanarity within the facets.
}
\item[\textbullet]{Generic, spatially discrete, time-frequency-baseline-direction-dependent 
full polarisation Jones matrices can be taken into account in the imaging and deconvolution
steps\footnote{\OMSOK{Is this really true? A-proj does support time-dependent Jones matrices.
And Wideband A-proj, supposedly, freq-dependent ones.}
This is not the case for
  \citet{Bhatnagar08,Tasse13} that are using
  A-projection and that need (i) spatially smooth DDE effects, and
  (ii) time-frequency-baseline independent Jones matrices for the PSF
  to be direction-independent.}.}
\item[\textbullet]{As shown in \Sec/ \ref{sec:DDEFacetting_IMCORR} and
  \ref{sec:DDEFacetting_DDPSF}, to account for time-frequency-baseline dependent Mueller matrices, we compute a direction
  dependent PSF for use in the minor cycle of deconvolution.}
\item[\textbullet] The above also allows for the effects of time and bandwidth
averaging to be explicitly incorporated into deconvolution.

\end{itemize}

In order to accurately deal with large fractional bandwidth of modern interferometers, one has to estimate the spectral properties of the
sky term in the Measurement Equation. We have implemented two new wideband deconvolution algorithms:

\begin{itemize}
\item[\textbullet]{A hybrid matching pursuit
  algorithm (\PMP/, see \Sec/ \ref{sec:PMPClean}), with similarities to \MTMSCLEAN/ \citep{Rau11}.}
\item[\textbullet]{A hybrid joint deconvolution
  algorithm that we call Subspace Deconvolution (\SSD/, see \Sec/ \ref{sec:SecSSD}). This does joint
  deconvolution on subsets of pixels (islands), while treating the islands independently. 
  For the former, we have implemented a genetic algorithm to perform the nonlinear optimisation step
  (\Sec/ \ref{sec:SSD_GA})}
\end{itemize}

Finally, a few interesting and innovative technical features are incorporated:

\begin{itemize}
\item[\textbullet] A general tessellation scheme that supports both regular (square) and nonregular (Voronoi 
tesselation) facets.
\item[\textbullet] On-the-fly baseline-dependent averaging within the gridder and degridder, resulting in a factor 
of several speedup of these operations.
\item[\textbullet] A sparsification scheme that dramatically accelerates the initial (shallow) major cycles,
when using a CLEAN-style deconvolution loop.
\item[\textbullet] Completely asynchronous multiprocessing, which does I/O and computation concurrently, and achieves
excellent scaling properties on NUMA architectures.
\end{itemize}

The current version of \DDFacet/ can deal with full polarisation
deconvolution, and can take into account externally defined
Jones matrices and/or beam patterns. It has been successfully tested
with data diverse telescopes such as LOFAR, VLA, MeerKAT AR1 and ATCA. 

Further development of \DDFacet/ is proceeding in a number of directions. These include
(i) extending the basis set of \SSDGA/ and \PMP/ with Gaussian components for
better deconvolution of extended emission, (ii) implementing other types of \SSD/ algorithms,
using convex solvers for the optimisation steps, (iii) implementing CS-derived minor cycle
algorithms, (iv) back-porting existing multiscale CLEAN algorithms into the framework. On the more 
technical side, we are working on incorporating (v) GPU-based gridders
and degridders, and (vi) the accelerated Direct Fourier Transform implemented in
\Montblanc/ \citep{Perkins15} for high dynamic range imaging. Finally, a distributed 
implementation of the framework is being designed.

\paragraph{Acknowledgments.} The research of
O. Smirnov is supported by the South African Research Chairs
Initiative of the Department of Science and Technology and National
Research Foundation. MJH acknowledges support from the UK Science and Technology
Facilities Council [ST/M001008/1]; some testing of \DDFacet/ was
carries out using the LOFAR-UK computing
facility located at the University of Hertfordshire and supported by
STFC [ST/P000096/1]. TWS acknowledges support from the ERC Advanced
Investigator program NewClusters 321271.

\bibliographystyle{aa}

\bibliography{references}



\end{document}